\documentclass[onecolumn,journal,11pt]{IEEEtran}
\usepackage{amsmath,amsthm,amssymb}
\usepackage{enumerate}
\usepackage{color}
\usepackage{xcolor}
\usepackage{url}
\usepackage{balance}
\usepackage{graphicx} 
\usepackage{mathrsfs}
\usepackage{epsfig}
\usepackage{verbatim}
\usepackage{setspace}
\usepackage{cite}
\usepackage{multicol}

\newtheorem{theorem}{Theorem}%[section]
\newtheorem{lemma}[theorem]{Lemma}

\newtheorem{claim}[theorem]{Claim}

\DeclareMathOperator*{\argmin}{arg\,min}
\DeclareMathOperator*{\argmax}{arg\,max}

\newcommand{\dnot}{\delta_0}

\newcommand{\dfour}{\delta_4}
\newcommand{\dfive}{\delta_5}

\newcommand{\dfourd}{\delta_4(\delta)}

\newcommand{\gammad}{\gamma(\delta)}

\newcommand{\ftwo}{f_2(\delta, \epsilon)}

\newcommand{\tdftwo}{\td{f}_2(\delta, \epsilon)}

\renewcommand{\vec}[1]{\mathbf{#1}}
\newcommand{\bV}{\mathbf{V}}
\newcommand{\bX}{\mathbf{X}}
\newcommand{\bv}{\mathbf{v}}
\newcommand{\bx}{\mathbf{x}}
\newcommand{\bY}{\mathbf{Y}}
\newcommand{\bZ}{\mathbf{Z}}
\newcommand{\by}{\mathbf{y}}
\newcommand{\bz}{\mathbf{z}}
\newcommand{\bS}{\mathbf{S}}
\newcommand{\bs}{\mathbf{s}}

\newcommand{\bU}{\mathbf{U}}
\newcommand{\bu}{\mathbf{u}}

\newcommand{\btdv}{\mathbf{\td{v}}}
\newcommand{\btdx}{\mathbf{\td{x}}}
\newcommand{\btds}{\mathbf{\td{s}}}
\newcommand{\btdy}{\mathbf{\td{y}}}
\newcommand{\btdz}{\mathbf{\td{z}}}

\newcommand{\btdV}{\mathbf{\td{V}}}

\newcommand{\btdS}{\mathbf{\td{S}}}
\newcommand{\btdY}{\mathbf{\td{Y}}}
\newcommand{\btdZ}{\mathbf{\td{Z}}}

\newcommand{\bvD}{\mathbf{v}^{\Delta}}
\newcommand{\bxD}{\mathbf{x}^{\Delta}}
\newcommand{\bsD}{\mathbf{s}^{\Delta}}
\newcommand{\byD}{\mathbf{y}^{\Delta}}
\newcommand{\bzD}{\mathbf{z}^{\Delta}}

\newcommand{\bVD}{\mathbf{V}^{\Delta}}

\newcommand{\bSD}{\mathbf{S}^{\Delta}}
\newcommand{\bYD}{\mathbf{Y}^{\Delta}}
\newcommand{\bZD}{\mathbf{Z}^{\Delta}}

\newcommand{\bbP}{\mathbb{P}}

\newcommand{\cU}{{\cal U}}
\newcommand{\cT}{{\cal T}}
\newcommand{\cX}{{\cal X}}
\newcommand{\cY}{{\cal Y}}
\newcommand{\cZ}{{\cal Z}}
\newcommand{\cS}{{\cal S}}
\newcommand{\cC}{{\cal C}}
\newcommand{\cL}{{\cal L}}
\newcommand{\cB}{{\cal B}}

\newcommand{\cP}{{\cal P}}

\newcommand{\sQ}{\mathscr{Q}}
\newcommand{\sT}{\mathscr{T}}

\newcommand{\td}[1]{\tilde{#1}}

\newcommand{\removed}[1]{}

\newcommand{\ajblue}[1]{}
\title{Communication over an Arbitrarily Varying Channel under a State-Myopic Encoder }
\author{
  \IEEEauthorblockN{Amitalok J. Budkuley and Sidharth Jaggi\\ }
	  \IEEEauthorblockA{Department of Information Engineering\\
		The Chinese University of Hong Kong, Hong Kong\\
    Emails: \{amitalok, jaggi\}@ie.cuhk.edu.hk }
}
\begin{document}
\maketitle 
\interdisplaylinepenalty=2500
\interfootnotelinepenalty=10
\thispagestyle{empty}
\begin{abstract}
We study the problem of communication over a discrete arbitrarily varying channel (AVC) when a noisy version of the state is known non-causally at the encoder. The state is chosen by an adversary which knows the coding scheme. A \emph{state-myopic} encoder observes this state non-causally, though imperfectly, through a noisy discrete memoryless channel (DMC). We first characterize the capacity of this state-dependent channel when the encoder-decoder share randomness unknown to the adversary, i.e., the randomized coding capacity. Next, we show that when only the encoder is allowed to randomize, the capacity remains unchanged when positive. Interesting and well-known special cases of the state-myopic encoder model are also presented.
\end{abstract}
\section{Introduction}\label{sec:introduction}
\begin{figure}[!ht]
  \begin{center}
    \includegraphics[trim=0cm 17cm 0cm 8cm, scale=0.35]{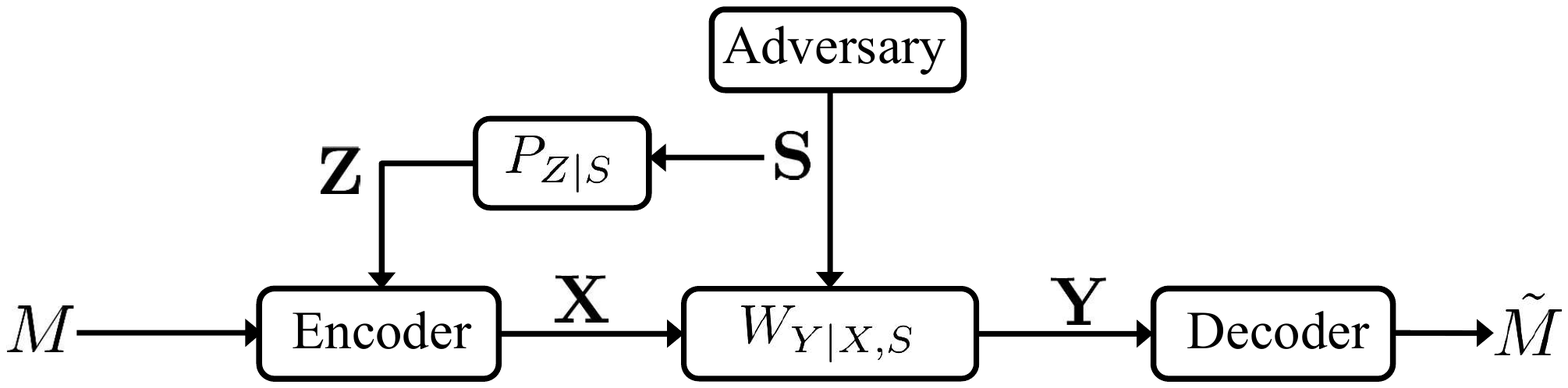}
    \caption{The communication setup}
    \label{fig:main:setup}
  \end{center}
\end{figure}

Consider the communication setup in Fig.~\ref{fig:main:setup}, where a message $M$ is sought to be transmitted over a memoryless state-dependent channel $W_{Y|X,S}$. 
The channel state is controlled by a jamming adversary which knows the coding scheme and can input arbitrary state vectors $\bS$, possibly  through randomized strategies. The adversary's choice of state $\bS$ is revealed non-causally, though  imperfectly, to the encoder. In particular, we assume that along with $M$, the encoder has a noisy or \emph{myopic} view\footnote{The myopic view model was introduced in~\cite{sarwate-itw2010}.} $\bZ$ of state $\bS$, where $\bS$ is observed through a  discrete memoryless channel (DMC) $P_{Z|S}$. In this work, we study the capacity of this state-dependent channel under a \emph{state-myopic} encoder.   

State-dependent channels, especially discrete channels which are the focus in the work, have received considerable attention in literature. For such channels, the capacity under non-causal awareness of the state at the encoder, when components of the \emph{random} state are generated according to a known independent and identically distributed (i.i.d.)  process, was characterized in the seminal work of Gel'fand and Pinsker~\cite{gelfand-pinsker} (see also~\cite{kusnetsov-tsybakov}). Cover and Chiang~\cite{cover-chiang-it2002} characterized the capacity when the encoder's non-causal view was corrupted via  a noisy DMC (see also~\cite{keshet-book2008} for a simpler proof). Versions of each of these problems under  causal knowledge of state have also appeared (cf.~\cite{shannon-ibm1958,salehi-1992,caire-shamai-it1999}). 

Ahlswede~\cite{ahlswede-it1986} analysed an \emph{adversarial} version of this \emph{Gel'fand-Pinsker problem}~\cite{gelfand-pinsker}, where the channel state $\bS$ may be chosen arbitrarily. 
%In that work, its channel capacity under deterministic coding was also characterized. 
Several other closely related adversarial channel models (see, for instance,~\cite{osullivan-moulin-ettinger-isit1998, cohen-lapidoth-it2002,bdp-it2017,pereg-isit2017}  and some of the references therein) have subsequently been studied. More generally, all of these channels belong to the class of arbitrarily varying channels (AVC), first proposed in~\cite{blackwell-ams1959}. The AVC framework has subsequently been  employed extensively to study varied adversarial communication problems. It is well known (cf.~\cite{csiszar-korner-book2011, lapidoth-narayan-it1998}) that the nature of results for AVCs crucially depend upon the assumptions made with respect to (w.r.t.) the communication system, for instance, the knowledge/capabilities possessed by the adversary and/or user. Adversary models have, in particular, received considerable attention. Several models have appeared, ranging from a `blind' or \emph{oblivious} adversary with no knowledge of the codeword~(e.g.~\cite{blackwell-ams1959,ahlswede-1978,csiszar-narayan-it1988-2}) to an \emph{omniscient} adversary with a perfect knowledge of the codeword~(e.g.~\cite{gilbert-bstj1952,langberg-focs2004}). More generally, a \emph{myopic} adversary with a noisy view  of the codeword was studied in~\cite{sarwate-itw2010} under randomized coding. 
A \emph{sufficiently myopic} adversary model, where the adversary's view is more noisy than the level of channel noise it can hope to induce,  was recently considered in~\cite{dey-isit2015}. 

In this work, reversing the gaze from the adversary to the user, we study the impact of myopicity at the encoder vis-\`{a}-vis the adversary's jamming state. %In particular, we assume that the encoder observes non-causally, a noisy version of the state vector. 
Our \emph{state-myopic} encoder model can be viewed as a bridge connecting Ahlswede's \emph{state-omniscient} model~\cite{ahlswede-it1986} with \emph{zero} or \emph{no} state-myopicity (i.e., under a full-rate observation channel) to the \emph{state-oblivious} model in~\cite{blackwell-ams1959,ahlswede-1978,csiszar-narayan-it1988-2} with \emph{full} state-myopicity (i.e., under a zero-rate observation channel). 
We refine this view through our main results.   
We first characterize the randomized coding capacity. Towards upper bounding the rate, our converse uses a memoryless, but crucially, a non-identically distributed jamming strategy which may depend on the code. Our proof of achievability uses the approach in~\cite{bdp-it2017}, and employs a \emph{refined} Markov lemma~\cite{bdp-it2017}. 
%Our achievability proof, which uses a \emph{refined}  Markov lemma~\cite{bdp-it2017}, directly addresses the randomized coding problem. 
This approach is different from the two-step approach in~\cite{ahlswede-it1986} which entails  first studying a \emph{compound channel} version (only memoryless jamming strategies permissible) of the problem,  followed by determining the randomized coding capacity using the `robustification technique'~\cite[pg.~625]{ahlswede-it1986}.
% for the achievability under randomized coding. 
We then  show that when only the encoder can privately randomize, the capacity remains unchanged when non-zero. 
%We also provide a brief exposition elaborating upon the aforementioned connections to well-known problems.
% 

The rest of the paper is organized as follows. We introduce the notation and the problem setup in Section~\ref{sec:problem}. The main results are stated in Section~\ref{sec:main:results}, while their proofs are presented in Section~\ref{sec:proofs}. We discuss some implications of our results, in particular, we elaborate upon aforementioned connections to well-known problems, and make concluding remarks in Section~\ref{sec:conclusion}.
\section{Notation and Problem Setup}\label{sec:problem}
\subsection{Notation} 
Let us denote random variables by upper case letters (e.g. $X$), the values they take by lower case letters (e.g. $x$) and their alphabets by calligraphic letters (e.g. $\mathcal{X}$). We use the boldface notation to denote random vectors (e.g. $\vec{X}$) and their values (e.g. $\vec{x}$). Here the vectors are of length $n$ (e.g. $\vec{X}=(X_1,X_2,\dots,X_n)$), where $n$ is the block length of operation. Let $\vec{X}^{i}=(X_1,X_2,\dots,X_i)$ and $\vec{x}^{i}=(x_1,x_2,\dots,x_i)$ as well as  $\bX_{i}^{k}=(X_i,X_{i+1},\dots,X_k)$ and $\bx_{i}^{k}=(x_i,x_{i+1},\dots,x_k)$. We use the $l_{\infty}$ norm denoted by $\|.\|_{\infty}$ for discrete vectors. For a set $\mathcal{X}$, let $\cP(\mathcal{X})$ be the set of all probability distributions on $\mathcal{X}$. Similarly, let us write as $\cP(\mathcal{X}|\mathcal{Y})$, the set of all conditional distributions of a random variable with alphabet $\mathcal{X}$ conditioned on another random variable with alphabet $\mathcal{Y}$. Let $X$ and $Y$ be two  random variables. Then, we denote the distribution of $X$ by $P_{X}(\cdot)$, the joint distribution of $(X,Y)$ by $P_{XY}(\cdot,\cdot)$ and the conditional distribution of $X$ given $Y$ by $P_{X|Y}(\cdot|\cdot)$. We denote the marginal distribution of $X$ obtained from $P_{X,Y}$ by $[P_{X,Y}]_X$.  Distributions corresponding to strategies adopted by the adversary are denoted by $Q$ instead of $P$ for clarity. 
%\ajb{In cases where the subscripts are clear from the context, we sometimes omit them to keep the notation simple. 
%For an event $E$, let $\mathbb{P}(E)$ denote the probability of $E$.} 
Functions will be denoted in lowercase letters (e.g., $f$).  
We denote a type of $X$ by $T_X$. Given sequences $\vec{x}$, $\vec{y}$, we
denote by $T_\vec{x}$ the type of $\vec{x}$, by $T_{\vec{x},\vec{y}}$ the joint
type of $(\vec{x},\vec{y})$ and by $T_{\vec{x}|\vec{y}}$ the conditional type
of $\vec{x}$ given $\vec{y}$. 
For $\epsilon\in(0,1)$, the set of $\epsilon$-typical  sequences $\vec{x}$ for a distribution $P_X$ is
%
%\begin{equation*}\label{eq:typicality}
$\mathcal{T}^{n}_{\epsilon}(P_X)=\{\vec{x}:\|T_\vec{x}-P_X\|_{\infty}\leq \epsilon\},$
%\end{equation*}
%
%where $\|.\|_{\infty}$ is the $l_{\infty}$ norm. 
and for a joint distribution $P_{X,Y}$ and $\vec{x}\in\cX^n$, the set of conditionally $\epsilon$-typical set of sequences $\vec{y}$, conditioned on $\vec{x}$, is defined as 
%
%\begin{equation*}
$\mathcal{T}^{n}_{\epsilon}(P_{X,Y}|\vec{x})=\{\vec{y}:\|T_{\vec{x},\vec{y}}-P_{X,Y}\|_{\infty}\leq \epsilon\}.$
%\end{equation*}
%
% 
\subsection{Problem Setup}
As shown in Fig.~\ref{fig:main:setup}, a message $M$ is sent over an AVC with user input $X$, jamming state $S$ and channel output $Y$. Random variables $X$, $S$ and $Y$ take values in finite sets $\cX$, $\cS$ and $\cY$ respectively. The channel behaviour is given by the fixed distribution $W_{Y|X,S}$. We consider the standard block coding framework with block length $n$, where $X_i$, $S_i$ and $Y_i$ denote the symbols associated with the $i$-th time instant. 
The jamming state $\bS$ is chosen by the adversary. Let $Q_{\bS}$ denote its distribution, which is arbitrary and unknown to user. A \emph{state-myopic} encoder receives two inputs: message $M$ and a noisy and non-causal version $\bZ$ of the state $\bS$. Here $\bZ$, where $Z_i\in\cZ$, $\forall i$ and $|\cZ|<\infty$, is output by a fixed DMC $P_{Z|S}$ under input $\bS$. The encoder  transmits $\bX$ on the channel. Upon receiving its noisy version $\bY$, the decoder outputs an estimate $\tilde{M}$ of the message $M$.

An $(n,R)$ \emph{deterministic code} of block length $n$ and rate $R$ 
%under non-causal noisy state knowledge $\bZ$ 
consists of a deterministic encoder-decoder pair $(\psi,\phi)$, where $\psi:\{1,2,\dots,2^{nR}\}\times
{\mathcal{Z}}^n\rightarrow {\mathcal{X}}^n$ and decoder $\phi:
{\mathcal{Y}}^n\rightarrow \{0,1,2,\dots,2^{nR}\}$, where an output of $0$ indicates decoding error. We  assume that $2^{nR}$ is an integer. 
\removed{
For an $(n,R)$ deterministic code, the maximum probability of error is given by
%
%\begin{IEEEeqnarray*}{rCl}\label{eq:pe:s:z}
$P_e^{(n)}:= \max_{m} \max_{Q_{\bS}} \bbP(\phi(\bY)\neq m|M=m),$
%&=& \frac{1}{2^{nR}} \sum_{i=1}^{2^{nR}} \left(\sum_{\by:\phi(\by)\neq i} \left( \prod_{i=1}^n W_{Y|X,S} (y_i|x_i,s_i)\right) \right)
%\end{IEEEeqnarray*}
%
where the probability is evaluated over the AVC $W_{Y|X,S}$, the channel $P_{Z|S}$,  and the adversary's action.\\
}%removed error definition for deterministic codes
An $(n,R)$ \emph{randomized code} of block length $n$ and rate $R$ is a random variable (denoted by $\Theta$) which takes values in the set of $(n,R)$ deterministic codes. 
For an $(n,R)$ randomized code, the maximum probability of error is given by
%
%\begin{IEEEeqnarray*}{rCl}\label{eq:pe:s:z}
$P_e^{(n)}:= \max_{m} \max_{Q_{\bS}} \bbP(\phi(\bY)\neq m|M=m),$
%&=& \frac{1}{2^{nR}} \sum_{i=1}^{2^{nR}} \left(\sum_{\by:\phi(\by)\neq i} \left( \prod_{i=1}^n W_{Y|X,S} (y_i|x_i,s_i)\right) \right)
%\end{IEEEeqnarray*}
%
where the probability is evaluated over the AVC $W_{Y|X,S}$, the channel $P_{Z|S}$, the shared randomness $\Theta$ and adversary's action. A rate $R$ is \emph{achievable} if for any $\epsilon>0$, there exists an $n_0(\epsilon)$ large enough such that for all $n\geq n_0(\epsilon)$ there exist $(n,R)$ randomized codes  with corresponding  $P_e^{(n)}$ less than $\epsilon$. We define the \emph{capacity} as the supremum of all achievable rates.
An $(n,R)$ \emph{code with  stochastic encoder} of block length $n$ and rate $R$ consists of a stochastic encoder-deterministic decoder pair $(\Psi,\phi)$, where $\Psi:\{1,2,\dots, 2^{nR}\}\times \cZ^n \rightarrow \cP(\cX^n)$ and $\phi:\cY^n\rightarrow \{0,1,2,\dots,2^{nR}\}$. Here an output of $0$ indicates a decoding error.  For an $(n,R)$ code with  stochastic encoder, the maximum probability of error is given by
%
%\begin{IEEEeqnarray*}{rCl}\label{eq:pe:s:z}
$P_e^{(n)}:= \max_{m} \max_{Q_{\bS}} \bbP(\phi(\bY)\neq m|M=m).$
%&=& \frac{1}{2^{nR}} \sum_{i=1}^{2^{nR}} \left(\sum_{\by:\phi(\by)\neq i} \left( \prod_{i=1}^n W_{Y|X,S} (y_i|x_i,s_i)\right) \right)
%\end{IEEEeqnarray*}
%
Here the probability is evaluated over the AVC $W_{Y|X,S}$, the channel $P_{Z|S}$, the encoding map and the adversary's action. 
%A rate $R$ is \emph{achievable} if for any $\epsilon>0$, there exists an $n_0(\epsilon)$ large enough such that for all $n\geq n_0(\epsilon)$ there exists an $(n,R)$ deterministic code  with corresponding  $P_e^{(n)}$ less than $\epsilon$. We define the \emph{capacity} as the supremum of all achievable rates. 
The definitions of achievable rate and capacity under codes with stochastic encoder can be analogously stated as earlier.
\section{The Main Results}\label{sec:main:results}
We now present our main results.
%, viz., the randomized coding capacity, followed by the deterministic coding capacity for the AVC $W_{Y|X,S}$ under non-causal but noisy observation of state $\bS$. Interestingly, the randomized coding capacity (given in Theorem~\ref{thm:main:result:dmc:r}) equals the deterministic coding capacity (given in Theorem~\ref{thm:main:result:dmc:r}). 
Define the set $\cP^Q(\cZ):=\{P_Z\in\cP(\cZ): P_Z=[Q_S P_{Z|S}]_Z, Q_S\in\cP(\cS)\}$. Given $P_Z\in\cP^Q(\cZ)$ and some $Q_S\in\cP(\cS)$, where $[P_{Z|S}Q_S]_Z=P_Z$, and under  fixed distribution $P_{U|Z}$ and function $x:\cU\times\cZ\rightarrow \cX$, let $I(U;Y)-I(U;Z)$ denote the mutual information quantity evaluated under the joint distribution $Q_{S} P_{Z|S} P_{U|Z} \vec{1}_{\{X=x(U,Z)\}} W_{Y|X,S}$. Let $\cU$ denote the alphabet of $U$. We define\footnote{As $I(U;Y)-I(U;Z)$ is a  continuous function of these variables, where the latter take values over compact sets, the min-max-min exists.}
\begin{IEEEeqnarray}{rCl}\label{eq:C:star}
C^*:=\hspace{-2mm}\min_{P_{Z}\in\cP^Q(\cZ)} \max_{  P_{U|Z},\ x(\cdot,\cdot)  } \hspace{-2mm} \min_{\substack{Q_{S}:\\ [Q_S P_{Z|S}]_Z=P_Z}} \hspace{-2mm}I(U;Y)-I(U;Z),\IEEEeqnarraynumspace
\end{IEEEeqnarray}
where $P_{U|Z}\in\cP(\mathcal{U}|\cZ)$, $x:\cU \times \cZ\rightarrow \cX$ and $|\cU|\leq |\cX|^{|\cZ|}$. 
We now state our first result.
\begin{theorem}\label{thm:main:result:dmc:r}
The randomized coding capacity $C_r$ under maximum probability of error criterion is 
\begin{equation}
C_r=C^*.\label{eq:cap:r}
\end{equation}
\end{theorem}
The proof of this result is presented in Section~\ref{sec:proofs}. %
%The next result gives the capacity characterization under codes with  stochastic encoder. 
%
\begin{theorem}\label{thm:main:result:dmc:s}
The capacity $C_s$ for codes with  stochastic encoder equals the randomized coding capacity when positive.
%For codes with  stochastic encoder, the capacity $C_s$ under a  maximum probability of error criterion is 
%
%\begin{equation}
%C_s=C^*.\label{eq:cap:s}
%\end{equation}
\end{theorem}
The proof of this result can be found in Section~\ref{sec:proofs}. \\
\noindent \emph{Remarks:}\\
{\it
1	. 
%Similar to other AVC models, it is seen that the capacities under randomized coding and under codes with  stochastic encoder are equal (cf.~\cite{csiszar-korner-book2011}). 
Although our results are stated under a maximum (over messages) probability of error criterion, they continue to hold under an average (over messages) probability of error criterion as well. This is because while the achievability is proved under the maximum probability of error criterion, our converse is proved under an average probability error of criterion.\\
2. Our state-myopic encoder model generalizes, through the degree of myopicity, the fully state-myopic model~\cite{blackwell-ams1959} as well as the zero state-myopic model~\cite{ahlswede-it1986}. Refer the discussion in Section~\ref{sec:conclusion} for details. 
}%end \it
\section{Proofs}\label{sec:proofs}
\subsection{Proof of Theorem~\ref{thm:main:result:dmc:r}}\label{sec:proof:r}
We first present the converse followed by the achievability. 
\subsubsection{Converse}
\label{gpavc:converse}
Our proof for the converse considers an average probability of error criterion instead of the maximum probability of error criterion. 
For this stronger version\footnote{This is owing to the fact that a rate $R$ which is not achievable under an average probability of error criterion, is not achievable under the maximum average of error criterion. } of the converse, let the average probability of error be
\begin{equation*}\label{eq:avgpe}
P^{(n)}_e:=\frac{1}{2^{nR}} \sum_{m=1}^{2^{nR}} P^{(n)}_{e,m},
\end{equation*}
where
\begin{equation*}
P^{(n)}_{e,m}=\max_{Q_{\vec{S}}} \mathbb{P}\left( \Phi(\vec{Y})\neq m|M=m\right).
\end{equation*}
Our converse will consider a specific memoryless, but possibly non-identically distributed, jamming strategy which depends on the randomized code. Under such a jamming strategy, we will upper bound the rate of reliable communication possible. 

Our proof of the converse starts along the lines of the converse for the standard Gel'fand-Pinsker problem~\cite{elgamal-kim}. Consider any sequence of randomized $(n,R)$ codes with average probability of error $P_e^{(n)}\rightarrow 0$  as $n\rightarrow \infty$. From Fano's inequality, we know that for such a sequence of codes $H(M|\bY,\Theta)\leq n\epsilon_n$, where $\epsilon_n\rightarrow 0$ as $n\rightarrow \infty$. 
\begin{IEEEeqnarray*}{rCl}
nR&\leq&  H(M)\\
&=&  I(M;\bY,\Theta)+H(M|\bY,\Theta)\\
&\leq& I(M;\bY,\Theta)+n\epsilon_n\\
&=& I(M;\Theta)+I(M;\bY|\Theta)+n\epsilon_n\\
&\stackrel{(a)}{=}& \sum_{i=1}^n I(M;Y_i|\bY^{i-1},\Theta)+n\epsilon_n\\
&\leq& \sum_{i=1}^n I(M,\bY^{i-1},\Theta;Y_i)+n\epsilon_n\\
&=& \sum_{i=1}^n I(M,\bZ_{i+1}^n,\bY^{i-1},\Theta;Y_i)-\sum_{i=1}^n I(Y_i;Z_{i+1}^n|M,\bY^{i-1},\Theta)+n\epsilon_n\\
&\stackrel{(b)}{=}& \sum_{i=1}^n I(M,\bZ_{i+1}^n,\bY^{i-1},\Theta;Y_i)-\sum_{i=1}^n I(\bY^{i-1};Z_i|M,\bZ_{i+1}^n\Theta,)+n\epsilon_n\\
&\stackrel{(c)}{=}& \sum_{i=1}^n I(M,\bZ_{i+1}^n,\bY^{i-1},\Theta;Y_i)-\sum_{i=1}^n I(M,\bZ_{i+1}^n,\bY^{i-1},\Theta;Z_i)+n\epsilon_n\\
&\stackrel{(d)}{=}& \sum_{i=1}^n I(U_i;Y_i)-\sum_{i=1}^n I(U_i;Z_i)+n\epsilon_n \\
&\stackrel{}{= }& n \left(\sum_{i=1}^n \frac{1}{n}\left(I(U_i;Y_i)-I(U_i;Z_i)\right) \right)+n\epsilon_n. \yesnumber\label{eq:conv1}
\end{IEEEeqnarray*}
%
%where $P_{U,X|Z}\in\cP(\cU\times\cX|\cZ)$ and $x:\cU\times \cZ\rightarrow \cX$. 
We get $(a)$ via the independence of $M$ and $\Theta$, while $(b)$ follows from Csisz\'{a}r's sum identity~\cite[pg.~25]{elgamal-kim}.  Under memoryless jamming,  $Z_i$ is independent of $(M,\bZ_{i+1}^n,\Theta)$, $\forall i$, which gives us $(c)$. By introducing the auxiliary variable $U_i:=(M,\bZ_{i+1}^n,\bY^{i-1},\Theta)$, $\forall i$, we have $(d)$.  

Although the adversary can employ arbitrary jamming strategies $Q_{\bS}$, we analyse the rate performance under memoryless jamming strategies. In particular, given a randomized encoding map, let us assume that the adversary restricts to only memoryless (though, possibly non-identically distributed) jamming strategies of the form
\begin{IEEEeqnarray*}{rCl}
Q_{\bS}(\bs):=\prod_{i=1}^n Q_{S_i}(s_i),
\end{IEEEeqnarray*}
where $Q_{S_i}\in\cP(\cS)$. Observe that under such memoryless jamming strategies, we have $(U_i,Z_i)\rightarrow (X_i,S_i)\rightarrow Y_i$, $\forall i$. Prior to specifying the jamming strategy $Q_{\bS}$, by specifying $Q_{S_i}(s_i)$ for each $ i$,  note that
% given $\{Q_{S_j}(s_j)\}_{j=1}^{i-1}$, 
we have
\begin{IEEEeqnarray*}{rCl}
&P_{U_i,X_i|Z_i}&(U_i=u_i,X_i=x_i|Z_i=z_i)\\
&\stackrel{(a)}{=}& P_{(M,\bZ_{i+1}^n,\bY^{i-1},\Theta), X_i|Z_i}((m,\bz_{i+1}^n,\by^{i-1},\theta),x_i|s_i)\\
&\stackrel{}{=}& \sum_{\bs^{i-1},\bx^{i-1},\bx_{i+1}^n,\bz^{i-1}} P_{(M,\bZ_{i+1}^n,\bY^{i-1},\Theta),\bS^{i-1},\bX^{i-1},\bX_{i+1}^n,\bZ^{i-1}, X_i|Z_i}((m,\bz_{i+1}^n,\by^{i-1},\theta),\bs^{i-1},\bx^{i-1},\bx_{i+1}^n,\bz^{i-1},x_i|z_i)\\
&\stackrel{}{=}& \sum_{\bs^{i-1},\bx^{i-1},\bx_{i+1}^n,\bz^{i-1}} P_{M}(m)  P_{\Theta}(\theta) P_{\bS^{i-1},\bZ^{i-1}}(\bs^{i-1},\bz^{i-1}) P_{\bZ_{i+1}^n}(\bz_{i+1}^n)  \\
&&\hspace{25mm} \cdot P_{(\bX^{i-1},X_i,\bX_{i+1}^n)|M,\Theta,\bZ} ((\bx^{i-1},x_i,\bx_{i+1}^n)|m,\theta,\bz) P_{\bY^{i-1}|\bX^{i-1},\bS^{i-1}} (\by^{i-1}|\bx^{i-1},\bs^{i-1}) \\
&\stackrel{}{=}& \sum_{\bs^{i-1},\bx^{i-1},\bx_{i+1}^n,\bz^{i-1}} P_{M}(m)  P_{\Theta}(\theta) \left[Q_{\bS^{i-1}}(\bs^{i-1}) P_{\bZ^{i-1}|\bS^{i-1}}(\bz^{i-1}|\bs^{i-1})\right]  \left[\sum_{\bs_{i+1}^{n}} Q_{\bS_{i+1}^{n}} (\bs_{i+1}^n )P_{\bZ_{i+1}^n|\bS_{i+1}^{n}}(\bz_{i+1}^n| \bs_{i+1}^n)  \right] \\
&&\hspace{25mm} \cdot P_{\bX|M,\Theta,\bZ} (\bx|m,\theta,\bz)  P_{\bY^{i-1}|\bX^{i-1},\bS^{i-1}} (\by^{i-1}|\bx^{i-1},\bs^{i-1})  \\
&\stackrel{}{=}& \sum_{\bs^{i-1},\bx^{i-1},\bx_{i+1}^n,\bz^{i-1}} P_{M}(m)  P_{\Theta}(\theta) \left[\prod_{j=1}^{i-1} Q_{S_j}(s_j) P_{Z|S}(z_j|s_j)\right]\left[\sum_{\bs_{i+1}^n} \left(\prod_{j=i+1}^n Q_{S_j}(s_j) P_{Z|S}(z_j|s_j)\right)\right] \\
&&\hspace{25mm} \cdot   P_{\bX|M,\Theta,\bZ} (\bx|m,\theta,\bz) \left[ \prod_{j=1}^{i-1} W_{Y|X,S} (y_j|x_j,s_j) \right].\yesnumber\label{eq:P:uxz}
\end{IEEEeqnarray*}
Here $(a)$ follows from substituting $U_i=(M,\bZ_{i+1}^n,\bY^{i-1},\Theta)$ and $u_i=(m,\bz_{i+1}^n,\by^{i-1},\theta)$. From~\eqref{eq:P:uxz}, it follows that $P_{U_i,X_i|Z_i}$ depends on $\{Q_{S_i}\}_{i=1}^n$. In particular, given $U_i=(M,\Theta,\bZ_{i+1}^n,\bY^{i-1})$, this dependence on the jamming strategy $Q_{\bS}$ is captured in a two-fold manner: dependence on all past outputs $\bY^{i-1}$ (depends on $Q_{\bS^{i-1}}$), and dependence on all future noisy observations $\bZ_{i+1}^n$ of the jamming state (depends on $Q_{\bS_{i+1}^n}$) at the encoder. We now curtail this dependence to only jamming strategies of the past  by further restricting the memoryless (possibly non-i.i.d.) jamming strategies $\{Q_{S_i}\}_{i=1}^n$  to those which induce a fixed i.i.d. marginal $P_Z\in\cP^Q(\cZ)$ at the encoder, i.e., $[Q_{S_i}P_{Z|S}]_Z=P_{Z_i}=P_{Z}$, $\forall i$, where $P_{Z}\in\cP^{Q}(\cZ)$. The effect of this restriction (cf.~\eqref{eq:P:uxz}) is that while $P_{U_i,X_i|Z_i}$  continues to depend on the randomized encoder $P_{\bX|M,\Theta,\bZ}$ in the same manner, it's dependence on the jamming strategy $Q_{\bS}$ is only via $\{Q_{S_j}\}_{j=1}^{i-1}$ and the known distribution $P_Z$. This allows us to now define $Q_{S_i}$ \emph{inductively} as follows.
For $i=1,2,\dots, n$, given $\{Q_{S_j}\}_{j=1}^{i-1}$ and $P_{U_i,X_i|Z_i}$, let 
\begin{IEEEeqnarray*}{rCl}
Q_{S_i}:= \argmin_{{\td{Q}}_{S_i}: [{\td{Q}}_{S_i} P_{Z|S}]=P_Z} (I(U_i;Y_i)-I(U_i;S_i)).\yesnumber\label{eq:min:Q:i}
\end{IEEEeqnarray*}
Thus, it follows from~\eqref{eq:conv1} and ~\eqref{eq:min:Q:i} that
\begin{IEEEeqnarray*}{rCl}
nR&\stackrel{}{\leq }& n \left(\sum_{i=1}^n \frac{1}{n} \min_{Q_{S_i}: [Q_{S_i} P_{Z|S}]_Z=P_Z } (I(U_i;Y_i)-\sum_{i=1}^n I(U_i;Z_i)) \right)+n\epsilon_n.    
%&\stackrel{}{\leq }& n \max_{P_{U,X|Z}} (I(U;Y)-I(U;Z))+n\epsilon_n,
\end{IEEEeqnarray*}
for $P_{U_i,X_i|Z_i}$, $i=1,2,\dots,n$. 
Further, observe that for $i=1,2,\ldots,n$,
\begin{align}\label{eq:U:countable}
\min_{Q_{S_i}: [Q_{S_i} P_{Z|S}]_Z=P_Z } \left(I(U_{i};Y_{i})- I(U_{i};Z_i)\right)\leq \max_{P_{U_i,X_i|Z_i}}\min_{Q_{S_i}: [Q_{S_i} P_{Z|S}]_Z=P_Z }\left(I(U_i;Y_i)- I(U_i;Z_i)\right), 
\end{align}
where the maximization is over all $P_{U_i,X_i|Z_i}$ 
with finite alphabet $\cU$ of $U_i$. Here the inequality in~\eqref{eq:U:countable} holds as the fixed $P_{U_i,X_i|Z_i}$ (induced by
the code) on the LHS is one such distribution.
Since the channel is memoryless, the RHS in~\eqref{eq:U:countable} does not depend on $i$, and thus, we have 
\begin{IEEEeqnarray*}{rCl}
R&\stackrel{}{\leq}& \max_{P_{U,X|Z}}\min_{Q_{S}:[Q_{S}P_{Z|S}]_Z=P_Z}\left(I(U;Y)- I(U;Z)\right)+\epsilon_n.
\end{IEEEeqnarray*}
Recall that we had restricted the adversary to memoryless jamming strategies $Q_{\bS}$, where $[Q_{S_i} P_{Z|S}]_Z=P_Z$, $\forall i$. Removing this restriction and noting that the adversary can choose any $Q_S\in\cP(\cS)$ so as to induce $P_{Z}\in\cP^{Q}(\cZ)\subseteq\cP(\cZ)$, we get
\begin{IEEEeqnarray*}{rCl}
R&\stackrel{}{\leq}&  \min_{P_{Z}\in\cP^Q(\cZ)} \max_{P_{U,X|Z}}\min_{Q_{S}:[Q_{S}P_{Z|S}]_Z=P_Z}\left(I(U;Y)- I(U;Z)\right)+\epsilon_n.   
\end{IEEEeqnarray*}
As this holds for all $n$, and we have $\epsilon_n\rightarrow 0$ as $n\rightarrow \infty$, it follows that
%The proof of the converse then follows by taking $n\rightarrow \infty$. 
\begin{IEEEeqnarray}{rCl}\label{eq:u:no}	
R&\stackrel{}{\leq}& \min_{P_Z \in\cP^Q(\cZ)}\max_{P_{U,X|Z}} \min_{Q_{S}:[Q_{S}P_{Z|S}]_Z=P_Z} \left(I(U;Y)- I(U;Z)\right).
\end{IEEEeqnarray}

We now show that instead of maximizing over $P_{U,X|Z}$, it is sufficient that the maximization in~\eqref{eq:u:no} is over distributions $P_{U|Z}$ and  functions $x:\cU\times \cZ\rightarrow \cX$, i.e., 
\begin{IEEEeqnarray}{rCl}\label{eq:max}
\min_{P_Z\in\cP^{Q}(\cZ)}\max_{P_{U,Z|X}}\min_{\substack{Q_{S}\\:[Q_{S} P_{Z|S}]=P_Z}} \left(I(U;Y)- I(U;Z)\right)=\min_{P_Z\in\cP^{Q}(\cZ)} \max_{P_{U|Z},\ x(\cdot,\cdot)}\min_{\substack{Q_{S}\\:[Q_{S}P_{Z|S}]_Z=P_Z}}\left(I(U;Y)- I(U;Z)\right).\IEEEeqnarraynumspace
\end{IEEEeqnarray}
Let us fix the optimizing distribution in $\cP^Q(\cZ)$, say $P^*_Z$, and the corresponding conditional distribution $P_{U|Z}$. The functional representation lemma~\cite[pg.~626]{elgamal-kim} guarantees the existence of a random variable $W$, independent of $(U,Z)$, such that $X$ is a function of $(W,U,Z)$.  Let $U'=(U,W)$ and let its alphabet be denoted by $\mathcal{U}'$. Then, we have $P_{U'|Z}((u,w)|z)=P_{U|Z}(u|z)P_W(w)$. Let the function be denoted by $x:\mathcal{U}'\times\mathcal{Z}\rightarrow\mathcal{X}$. Note that, as required, $(U',Z)\rightarrow (X,S)\rightarrow Y$ is a Markov chain. We now calculate the mutual information quantities under $U'$. 
\begin{IEEEeqnarray*}{rCl}
I(U';Z)&=&I(U,W;Z)\\
&=& I(U;Z)+I(W;Z|U)\\
&=& I(U;Z),\yesnumber\label{eq:us}
\end{IEEEeqnarray*}
where $W\perp\!\!\!\perp (U,Z)$ gives us the last inequality. Further, for any $Q_{S}\in\cP(\cS)$ such that $[Q_{S} P_{Z|S}]_Z=P_{Z}^*$,
\begin{IEEEeqnarray*}{rCl}
I(U';Y)&=&I(U,W;Y)\\
&=& I(U;Y)+I(W;Y|U)\\
&\geq & I(U;Y),
\end{IEEEeqnarray*}
and hence, 
\begin{IEEEeqnarray*}{rCl}
\min_{Q_{S}:[Q_{S} P_{Z|S}]_Z=P_{Z}^*} I(U';Y)\geq  \min_{[Q_{S} P_{Z|S}]_Z=P_{Z}^*}I(U;Y).\yesnumber\label{eq:uy}
\end{IEEEeqnarray*}
It then follows from~\eqref{eq:us} and~\eqref{eq:uy} that for the minimizing distribution $P_{Z}^*$, we have 
\begin{IEEEeqnarray*}{rCl}
\min_{[Q_{S} P_{Z|S}]_Z=P_{Z}^*} I(U;Y)-I(U;Z)&\leq& \min_{[Q_{S} P_{Z|S}]_Z=P_{Z}^*} I(U';Y)-I(U';Z).\label{eq:u:u'}
\end{IEEEeqnarray*}
Note that the LHS above is evaluated under a  conditional distribution $P_{X|U,Z}$ and the RHS under the corresponding $P_{U'|Z}$ and $x:\mathcal{U}'\times\mathcal{Z}\rightarrow\mathcal{X}$.
As the inequality holds for any $P_{X|U,Z}$, we have~\eqref{eq:max}, and thus
\begin{IEEEeqnarray}{rCl}\label{eq:R:1}
R&\stackrel{}{\leq}& \min_{P_Z\in\cP^{Q}(\cZ)}\max_{P_{U|Z},\ x(u,z)}\min_{\substack{Q_{S}\\:[Q_S P_{Z|S}]_Z=P_Z}}\left(I(U;Y)- I(U;Z)\right).
\end{IEEEeqnarray}

Finally, the \emph{Shannon strategy} approach (see, for instance,~\cite[Remark~7.6]{elgamal-kim}) gives us the bound on the cardinality of $|\cU|$. The maximization over
functions $x(u,z)$ in~\eqref{eq:R:1} can be equivalently viewed as a
maximization over the set of all functions $x_u:\cZ\rightarrow \cX; u\in \cU$.
As exactly $|\cX|^{|\cZ|}$
such distinct functions exist, without loss of generality, we can restrict $\cU$ to be of cardinality
at most $|\cX|^{|\cZ|}$.
This completes the proof of the converse. 
\subsubsection{Achievability}
We provide an outline of the proof of achievability. Our outline includes a brief description of the randomized code followed by an overview of the error analysis. The detailed proof uses the approach in~\cite{bdp-isit2017} and can be found in Appendix~\ref{app:ach:r}.\\
\emph{Code design:}
\begin{itemize}
\item For every type $T_Z\in\cP^{Q}(\cZ)$, choose  the optimal $P_{U|Z}$ and $x(\cdot,\cdot)$ according to~\eqref{eq:cap:r}.
Now we generate $2^{nR_U (T_Z)}$ i.i.d. $\bU$ sequences, each i.i.d. $\sim P_U$, where $P_U := \left[T_Z P_{U|Z}\right]_U$, to form the codebook $\cC(T_Z)$.   Each codebook $\cC(T_Z)$ is randomly partitioned into $2^{n(R_U (T_Z)-\tilde{R}(T_Z))}$ bins. 
%The exact values for $R_U(T_Z),\td{R}(T_Z)>0$ will be specified later.
Fix $\epsilon>0$ and define\footnote{The notation $P'_Z\stackrel{f(\epsilon)}{\approx} P''_Z$ means that for distributions $P'_Z,P''_Z\in\cP(\cZ)$, we have $\|P'_Z-P''_Z\|_\infty\leq f(\epsilon)$, where $f(\epsilon)>0$ and $f(\epsilon)\rightarrow 0$ as $\epsilon\rightarrow 0$.}
\begin{IEEEeqnarray*}{rCl}
R_{U}(T_Z)&:=&\min_{ \substack{ T_S: [T_S P_{Z|S}]_Z \stackrel{f(\epsilon)}{\approx} T_Z}} I(U;Y)-\epsilon  \nonumber\label{eq:R:U:tz}\\
\td{R}(T_Z)&:=&I(U;Z)+\epsilon  \nonumber\label{eq:R:td:tz}\\ 
R(T_Z)&:=& R_{U}(T_Z)-\td{R}(T_Z),\yesnumber\label{eq:R:tz}
\end{IEEEeqnarray*}
\item Our randomly generated code contains this list of binned codebooks for every $T_Z\in\cP^{Q}(\cZ)$, and is shared between the encoder and decoder. Through the available shared randomness $\Theta$, the encoder-decoder will jointly select one code from this ensemble and use it for communication. This process is equivalent to the code being randomly generated and then shared between the encoder and the decoder.
\end{itemize}
\emph{Encoder operations:}
\begin{itemize}
\item The encoder knows $m$ and $\bz$. It first calculates the type $T_{\bz}$. Next, it identifies codebook $\cC(T_{\bz})$ and the corresponding optimal pair $(P_{U|Z},x(u,z))$ (via~\eqref{eq:cap:r}). Within bin $m$ of the codebook $\cC(T_{\bz})$, it now checks to see if there exists any codeword $\bu_{m,k}$, $k=1,2,\dots, 2^{n\td{R}(T_{\bz})}$, jointly typical with the observed $\bz$ under the distribution $P_{U|Z} T_{\bz}$. If so, let the chosen codeword be $\bu$, else let $\bu:=\bu_{1,1}$. 
\item Next, it generates $\bx$, where $x_i=x(u_i,z_i)$, $i=1,2,\dots, n$. It then sends the type $T_{\bz}$\footnote{This is possible with negligible rate overhead and vanishing error probability when capacity is non-zero.} and $\bx$ over the channel. As there are up to a polynomial number of types~\cite{csiszar-korner-book2011}, for large enough $n$, the rate required to convey $T_{\bz}$ is at most
$\epsilon/2$. 
\item Observe that the overall rate of this coding scheme (message rate is given by the smallest rate of codebook $\cC(T_Z)$ for any type $T_Z\in\cP^Q(\cZ)$)
\begin{align*}
R 
& \leq \min_{T_Z\in\cP^Q(\cZ)} R (T_Z) +\epsilon/2\\
& \stackrel{(a)}{=} \min_{T_Z\in\cP^Q(\cZ)} (R_U (T_Z)-\tilde{R}(T_Z)-2\epsilon) +\epsilon/2\\
& \stackrel{}{\leq} \min_{P_Z\in\cP^Q(\cZ)} \max_{P_{U|Z},x (.,.)}\min_{\substack{Q_S\\ [Q_S P_{Z|S}]_Z=P_Z}} \hspace{-3mm} I(U;Y)-I(U;Z)\\ %{Q_S\in\sQ(P_Z)}
&\hspace{5mm} -2\epsilon +\epsilon/2+\epsilon/2\\
& \stackrel{}{=} C-\epsilon.
%& \leq \min_{P_Z\in\cP^Q(\cZ)} \max_{P_{U,X|Z}} \min_{Q_S}
%& \leq \max_{P_Y^\prime \in \cP(\cY)} \left[I_{P_Y^\prime}(U;Y)- \min_{\begin{array}{c}Q_{J|X}\in \sQ \\ P_Y \stackrel{f(\delta)}{\approx} P_Y^\prime\end{array}}I_{Q_{J|X}}(U;Z)\right] + \frac{\epsilon}{2} + \frac{\epsilon}{4} \quad \text{(using \eqref{eq:ru} and \eqref{eq:rtilde})}\\
%& \leq R^{(P_{U|Y},\tilde{x})}+\epsilon. \hspace{10mm} \text{(using \eqref{eq:rds},~\eqref{eq:ru} and~\eqref{eq:rtilde})}
\end{align*}
where $(a)$ follows from~\eqref{eq:R:tz}.
\end{itemize}
\emph{Decoder operations:}
\begin{itemize}
\item The decoder knows $T_{\bz}$ and observes channel output $\by$. It first identifies the set of conditional types
\begin{align*}
\sQ^{(n)}(T_\bz) & := \{T_{S}\in \cP(\cS) : [T_S P_{Z|S}]_Z \hspace{0mm} \stackrel{f(\epsilon)}{\approx} T_\bz\}.
\end{align*}
%where $f(\epsilon)>0$, and $f(\epsilon)\rightarrow 0$ as $\epsilon\rightarrow 0$.
The set $\sQ^{(n)}(T_\bz)$ contains types $T_S$ which result in a $Z$-marginal distribution which is close to the observed type $T_{\bz}$.
\item 
The decoder next determines the set of codewords $\bu$ such that $(\bu,\by)$ are jointly typical w.r.t. the distribution 
$\left[T_S P_{Z|S} P_{U|Z} \vec{1}_{\{X=x(U,Z)\}} W_{Y|X,S}\right]_{U,Y}$ for some type $T_{S} \in \sQ^{(n)}(T_{\bz})$. 
If there is a unique such codeword $\bu$, then it  outputs its bin index as the message estimate. Otherwise, it outputs $\td{m}=0$ indicating decoding error.
\end{itemize}
\noindent \emph{Error analysis:}\\
An error can occur for actual codeword $\bu\in\cC(T_{\bz})$ due to (i) $\bu$ not being decoded correctly, and (ii) some wrong codeword $\bu'\in\cC(T_{\bz})$ decoded incorrectly. A decoding error for the actual codeword can occur under the following cases:
\begin{itemize}
\item Given $m$ and $\bz$, the encoder cannot find within codebook $\cC(T_{\bz})$ any $\bu$ jointly typical with $\bz$ w.r.t. the joint distribution $P_{U|Z} T_{\bz}$. However, as the rate $\td{R}(T_Z)>I_{T_{\bz}}(U;Z)$, the probability that the encoder cannot find such a codeword $\bu$ is exponentially small (via covering lemma~\cite{elgamal-kim}).
\item Let $\bu$ be the codeword chosen by the encoder. A decoding error can occur if this $\bu$ does not satisfy the decoding condition under any jamming state $\bs$. We show that with high probability (w.h.p.)~\footnote{All our w.h.p. statements
hold under ``except for an exponentially small probability.''} such a possibility is precluded. 
Note that $\bz$ observed is typical w.r.t $P_{Z|S} T_{\bs}$. In fact, the type $T_{\bz}\in\cP^Q(\cZ)$ and $T_{\bz}$ is `close to' the Z-marginal $[P_{Z|S} T_{\bs}]_Z$. This implies that $T_{\bs}$ is one of the types considered by the decoder, i.e., $T_{\bs}\in\sQ(T_{\bz})$. As $(\bu,\bz)$ are jointly typical according to $P_{U|Z}T_{\bz}$, they are also jointly typical (though with a slightly larger slack) w.r.t. $P_{U|Z} [P_{Z|S} T_{\bs}]_Z$. We now use a version of the refined Markov lemma~\cite[Lemma~8]{bdp-it2017} to show that $(\bu,\bz,\bs)$ are jointly typical according to $P_{U|Z} P_{Z|S} T_{\bs}$. As $\bx$ is generated via function $x(U,Z)$, it follows (using  a version of the conditional typicality lemma~\cite{elgamal-kim}) that $(\bu,\bz,\bs,\bx)$ are jointly typical according to $P_{U|Z} P_{Z|S} T_{\bs} \vec{1}_{\{X=x(U,Z)\}}$. A similar argument guarantees that $\by$ generated through the memoryless channel $W_{Y|X,S}$ is such that the tuple $(\bu,\bz,\bs,\bx,\by)$ is w.h.p. jointly typical according to $P_{U|Z}P_{Z|S} T_{\bs} \vec{1}_{\{X=x(U,Z)\}} W_{Y|X,S}$. Thus, it follows that w.h.p. $(\bu,\by)$ are jointly typical according to the distribution $[P_{U|Z}P_{Z|S} T_{\bs} \vec{1}_{\{X=x(U,Z)\}} W_{Y|X,S}]_{U,Y}$. This guarantees that w.h.p. the actual codeword will be decoded correctly  at the decoder.
\end{itemize}
For a decoding error possibly caused by wrong codewords:
\begin{itemize}
\item The decoder receives the type $T_{\bz}$  and identifies the codebook $\cC(T_{\bz})$. Owing to our choice of $R_U(T_{\bz})$, for any given `candidate' $T_{S}\in \sQ(T_{\bz})$, the probability that there exists some codeword $\bu'$ jointly typical with $\by$ w.r.t. the distribution $[T_S P_{Z|S} P_{U|Z}\vec{1}_{\{X=x(U,Z) W_{Y|X,S}\}}]_{U,Y}$ is exponentially small (via the packing lemma~\cite{elgamal-kim}). As there are only up to a polynomial number of types $T_S$, the probability that the above error event can occur for any $T_{S}\in\sQ(T_{\bz})$ is also exponentially small. 
\end{itemize}
%\ajb{AJB: overview ends. detailed presentation and analysis follow  now.}\\
%
\subsection{Proof of Theorem~\ref{thm:main:result:dmc:s}}\label{sec:proof:d}
We have already established a more general converse under randomized coding in Theorem~\ref{thm:main:result:dmc:r}. Our proof of achievability  uses an approach similar to that in~\cite{ahlswede-it1986} and has two parts:
\begin{itemize}
\item[(a)] de-randomization: to show that a shared randomness of $O(2\log(n))$ bits is sufficient to achieve $C_r$.
\item[(b)] code concatenation: to show that there exists (under non-zero capacity) a concatenated code with  stochastic encoder which achieves randomized coding capacity.
%\item[(c)] capacity under deterministic coding: using the stochastic encoding scheme, we construct a deterministic coding scheme which achieves capacity.
\end{itemize}
\emph{Part (a):} Recall that in Theorem~\ref{thm:main:result:dmc:r}, we established the existence of a randomized code, say $\cC=(\Psi,\Phi)$, of any rate arbitrarily close to the capacity (capacity $C_r$ given in~\eqref{eq:cap:r}) with vanishing maximum probability of error. Thus, given any $\epsilon>0$, for the code $\cC$ we have $P_{e}^{(n)}\leq \epsilon$, i.e., $P_{e}^{(n)}(m,\bs)\leq \epsilon$, $\forall m, \bs$.
Consider $K$ independent repetitions of a random experiment of (deterministic) codebook selection from the randomized code $\cC$ (i.e, i.i.d. selection via  the randomized code distribution). Let  the $K$ outcomes be the deterministic codes $C_i:=(\psi_i,\phi_i)$, $i=1,2,\dots, K$, where $(\psi_i,\phi_i)$ denote the encoder-decoder  pair for code $C_i$. Given message $m$, jamming state $\bs$ and code $C_i$, let the resulting probability of error be $P_{e}^{(n)}(m,\bs, C_i)$, where the probability is over $W_{Y|X,S}$ and $P_{Z|S}$. Note that
\begin{IEEEeqnarray*}{rCl}
\mathbb{E}_{\cC}[P_{e}^{(n)}(m,\bs, C_i)]\leq \epsilon\,\,\,\,\forall m,\bs\yesnumber\label{eq:avg:tdD}.
\end{IEEEeqnarray*}
We now use Bernstein's trick~\cite{ahlswede-it1986} and note that for any $\mu>0$,
\begin{IEEEeqnarray*}{rCl}
\mathbb{P}_{\cC}\left(\frac{1}{K} \sum_{i=1}^K P_{e}^{(n)}(m,\bs, C_i) \geq \mu  \right)&=& \mathbb{P}_{\cC}\left( \sum_{i=1}^K P_{e}^{(n)}(m,\bs, C_i)\geq K \mu  \right) \\
&=&  \mathbb{P}_{\cC} \left( e^{  \sum_{i=1}^K P_{e}^{(n)}(m,\bs, C_i)}  \geq e^{ K\mu}\right) \\
&\stackrel{(a)}{\leq} & \exp(-K\mu) \mathbb{E}_{\cC}\left[  e^{ \sum_{i=1}^K P_{e}^{(n)}(m,\bs, C_i)  }  \right], \yesnumber\label{eq:P1}
\end{IEEEeqnarray*}
where $(a)$ follows from the Markov inequality. As $P_{e}^{(n)}(m,\bs, C_i)$, $i=1,2,\dots, K$, are independent and identically distributed,  it follows that
\begin{IEEEeqnarray*}{rCl}
\mathbb{E}_{\cC}\left[  e^{  \sum_{i=1}^K P_{e}^{(n)}(m,\bs, C_i) }  \right]&=& \left(\mathbb{E}_{\cC}\left[ e^{ P_{e}^{(n)}(m,\bs, C_i) } \right] \right)^K\\
&\stackrel{}{=}& \left(\mathbb{E}_{\cC}\left[1+\sum_{l=1}^{\infty} \frac{\left(P_{e}^{(n)}(m,\bs, C_i)\right)^l}{l!}\right]\right)^K\\
&\stackrel{(a)}{\leq}& \left(\mathbb{E}_{\cC}\left[1+\sum_{l=1}^{\infty} \frac{P_{e}^{(n)}(m,\bs, C_i)}{l!}\right]\right)^K\\
&\stackrel{}{=}& \left(1+\sum_{l=1}^{\infty} \frac{\mathbb{E}_{\cC}[P_{e}^{(n)}(m,\bs, C_i)]}{l!}\right)^K\\
&\stackrel{(b)}{\leq}& \left(1+\sum_{l=1}^{\infty} \frac{\epsilon}{l!}\right)^K\\
%&\stackrel{(d)}{\leq}& \left(1+\sum_{l=1}^{\infty} \frac{\td{D}^{(n)}}{l!}\right)^K\\
&\stackrel{}{=}& \left(1+\epsilon \sum_{l=1}^{\infty} \frac{1}{l!}\right)^K\\
&\stackrel{}{=}& \left(1+\epsilon (e-1)\right)^K\\
&\stackrel{}{\leq}& \left(1+\epsilon e\right)^K.\yesnumber\label{eq:D1}
\end{IEEEeqnarray*}
We get $(a)$ by noting that $P_{e}^{(n)}(m,\bs, C_i)\in[0,1]$, $\forall m,\bs,C_i$, while $(b)$ follows from~\eqref{eq:avg:tdD}.
Thus, from~\eqref{eq:P1} and~\eqref{eq:D1}, we get
\begin{IEEEeqnarray*}{rCl}
\mathbb{P}_{\cC}\left(\frac{1}{K} \sum_{i=1}^K P_{e}^{(n)}(m,\bs, C_i)\geq \mu  \right)&\stackrel{}{\leq} & e^{-K\mu} \left(1+\epsilon e\right)^K\\
&\stackrel{}{=} & e^{-K\mu} e^{K \log(1+\epsilon e)}\\
&=& e^{-K( \mu-\log(1+\epsilon e))}.
\end{IEEEeqnarray*}
Allowing  for any $m,\bs$, and taking the union bound, the maximum probability of error under the randomized code $(\Psi,\Phi):=\{(\psi_i,\phi_i)\}_{i=1}^{K}$ is
\begin{IEEEeqnarray*}{rCl}
\mathbb{P}_{\cC}\Bigg(\exists m,\bs: \frac{1}{K} \sum_{i=1}^K P_{e}^{(n)}(m,\bs, C_i)\geq \mu \Bigg)&\stackrel{}{\leq} & 2^{nR} |\cS|^n e^{-K( \mu-\log(1+\epsilon e))}\\
&=& e^{n(R\log(2)+\log|\cS|)} e^{-K( \mu-\log(1+\epsilon e))}\\
&=&  e^{-(K( \mu-\log(1+ \epsilon e))-n(R \log(2)+\log|\cS|))},
\end{IEEEeqnarray*}
which is vanishing as $n\rightarrow \infty$ when
\begin{IEEEeqnarray*}{rCl}
K>\frac{n(R\log(2)+\log|\cS|)}{\mu-\log(1+\epsilon e)},
\end{IEEEeqnarray*}
or, more simply, when $$K=n^2.$$ 
As $\epsilon>0$ is arbitrary, this %given a randomized code $\cC$ of arbitrary ensemble size that achieves the randomized coding capacity given in~\eqref{eq:cap:r}, 
implies that that for any rate $R<C_r$, there exists a randomized code with an ensemble comprising up to $n^2$ deterministic codebooks such that its  maximum probability of error is vanishing as $n\rightarrow \infty$.\\
%\footnote{Observe that this translates to a key rate $\frac{2\log n}{n}\rightarrow 0$ as $n\rightarrow \infty$. }
%This completes proof of this part.

\noindent \emph{Part (b):}
We now show that when the capacity is positive, there exists a code with  stochastic encoder which also achieves the randomized coding capacity. For any positive $R<C_r$ and any $\mu>0$, consider an $(n,R)$  randomized code $(\Psi,\Phi):=\{(\psi_i,\phi_i)\}_{i=1}^{n^2}$ with maximum probability of error $P_{e}^{(n)}\leq \mu$ (we know from part(a) that such a code exists). As the capacity is positive, it follows that there exists a simple coding scheme with a stochastic encoder $(\Psi^{\Delta},\phi^{\Delta})$ with block length $n^{\Delta}$ (here $n^{\Delta}=o(n)$), where $\Psi^{\Delta}:\{1,2,\dots,n^2\}\rightarrow \cP(\cX^{n^{\Delta}})$ and $\phi:\cY^{n^{\Delta}}\rightarrow \{1,2,\dots,n^2\}$, and its maximum probability of error $\td{\mu}<\mu$, such that the rate is arbitrarily small for large enough $n$. Using code $(\Psi^{\Delta},\phi^{\Delta})$  along with $(\Psi,\Phi)$, we now define a new concatenated code with  stochastic encoder $(\td{\Psi},\td{\phi})$ over a block length $\td{n}:=n+n^{\Delta}$. Given the larger block length $\td{n}$, let random vectors and their actual values be denoted by $\btdV:=(\bVD,\bV)$ and $\btdv:=(\bvD,\bv)$. 
For our concatenated code with  stochastic encoder, let $\td{\Psi}: \{1,2,\dots, 2^{nR}\}\times \cZ^{\td{n}}\rightarrow \cP(\cX^{\td{n}})$ and $\td{\phi}: \cY^{\td{n}}\rightarrow \{1,2,\dots,2^{nR}\}$ be given as
\begin{IEEEeqnarray}{rCl}
\td{\Psi}(m,\btdz)&:=&\frac{1}{n^2} \sum_{k=1}^{n^2} \Psi^{\Delta}(k)\times \psi_{k}(m,\bz)\label{eq:stoc:enc}\\
\td{\phi}(\btdy)&:=&\bigcup_{k=1}^{n^2} \phi^{\Delta}(\byD)\times \phi_k(\by).\label{eq:s:dec}
\end{IEEEeqnarray}
Given message $m$ and jamming state $\btds=(\bsD,\bs)$, we now evaluate the probability of \emph{correct decision} $P_{c}^{\td{n}}(m,\btds)$ under this code with stochastic encoder $(\td{\Psi},\td{\phi})$. 
\begin{IEEEeqnarray*}{rCl}
P^{(\td{n})}_{c}(m,\btds)&=& \sum_{\btdz} P_{\btdZ|\btdS}(\btdz|\btds) \sum_{\btdx} \td{\Psi}(\btdx|m,\btdz) \bbP_W(\td{\phi}(\btdY)=m|m,\btdz,\btdx,\btds)\\
&\stackrel{(a)}{=}& \sum_{\btdz} P_{\btdZ|\btdS}(\btdz|\btds) \sum_{\btdx=(\bxD,\btdx)} \left[\frac{1}{n^2}\sum_{k=1}^{n^2} \Psi^{\Delta}(\bxD|k) \times \vec{1}_{\{\bx=\psi_{k}(m,\bz)\}} \right] \\
&&\hspace{20mm}  \cdot \bbP_W \left(\bigcup_{k=1}^{n^2}  \Big\{ \{\phi^{\Delta}(\bYD)=k\}\times \{\phi_{k}(\bY)=m\} \Big\}\Big| m,\btdz,\btdx=(\bxD,\phi_k(m,\bz)),\btds \right)\\
&\stackrel{(b)}{\geq}& \frac{1}{n^2}\sum_{k=1}^{n^2} \sum_{\btdz} P_{\btdZ|\btdS}(\btdz|\btds) \sum_{\bxD} \Psi^{\Delta}(\bxD|k)   \\
&&\hspace{20mm}  \cdot \bbP_W \left(   \{\phi^{\Delta}(\bYD)=k\}\times \{\phi_{k}(\bY)=m\}\Big| m,\btdz,\btdx=(\bxD,\phi_k(m,\bz)),\btds \right)\\
&\stackrel{(c)}{=}& \frac{1}{n^2}\sum_{k=1}^{n^2} \left(\sum_{\bzD} P_{\bZD|\bSD}(\bzD|\bsD) \sum_{\bxD} \Psi^{\Delta}(\bxD|k) \bbP_W \left(   \{\phi^{\Delta}(\bYD)=k\}\Big| m,\bxD,\bsD\right) \right)\\
&&\hspace{20mm}  \cdot \left(\sum_{\bz}  P_{\bZ|\bS}(\bz|\bs) \bbP_W \left(   \{\phi_{k}(\bY)=m\}\Big| m,\bx=(\phi_k(m,\bz)) \right)  \right) \\
&\stackrel{(d)}{\geq}& \frac{1}{n^2}\sum_{k=1}^{n^2} \left(1-\td{\mu} \right) \left(\sum_{\bz}  P_{\bZ|\bS}(\bz|\bs) \bbP_W \left(   \{\phi_{k}(\bY)=m\}\Big| m,\bx=(\phi_k(m,\bz)) \right)  \right) \\
&= &  \left(1-\td{\mu} \right) \left[\frac{1}{n^2}\sum_{k=1}^{n^2}\left(\sum_{\bz}  P_{\bZ|\bS}(\bz|\bs) \bbP_W \left(   \{\phi_{k}(\bY)=m\}\Big| m,\bx=(\phi_k(m,\bz)) \right)  \right) \right] \\
&\stackrel{(e)}{\geq}&  \left(1-\td{\mu} \right) (1-\mu) \\
&\stackrel{(f)}{\geq}&  (1-\mu)^2 \\
&\stackrel{}{\geq}&  1-2\mu.
\end{IEEEeqnarray*}
Here $(a)$ follows from~\eqref{eq:stoc:enc} and~\eqref{eq:s:dec}, while $(b)$ follows from using the union bound w.r.t. the decoding event. We get $(c)$ from the fact that the channel is memoryless.
%as well as the nature of the encoder-decoder mappings given in~\eqref{eq:stoc:enc} and~\eqref{eq:s:dec}. 
As the code $(\Psi^{\Delta},\phi^{\Delta})$ has a maximum error probability of $\td{\mu}$, we get $(d)$. Noting that for the randomized code $(\Psi,\Phi)=\{\psi_k,\phi_k\}_{k=1}^{n^2}$, the probability of correct decision  $P_{c}^{(n)}:=1-P_{e}^{(n)}$, and thus, $P_{c}^{(n)}\geq 1-\mu$, we have $(e)$. Finally, we get $(f)$ as $0<\td{\mu}<\mu$. 
Thus, we have shown that for this code with a stochastic encoder $(\td{\Psi},\td{\phi})$, the probability of correct decision under every $m$ and $\btds$, is at least $(1-\mu)$, which directly implies that its  maximum probability of error is at most $\mu$. As the choice of  $\mu>0$ was arbitrary and since the rate penalty (due to $(\Psi^{\Delta},\phi^{\Delta})$) is vanishing as $n\rightarrow \infty$, the proof is complete.
	\section{Discussion and Conclusion}\label{sec:conclusion}
	Our \emph{state-myopic} encoder model unifies an entire spectrum of problems with encoder models ranging from the \emph{fully} state-myopic model~\cite{blackwell-ams1959} to the \emph{zero} state-myopic model~\cite{ahlswede-it1986} as discussed below.

	 \emph{Full State-Myopicity:}
	Owing to full myopicity, the encoder observes $Z\perp S$, where $Z$ has a fixed distribution  $P_Z$. Thus, the encoder learns nothing about the state $S$, and hence, disregards $Z$. This makes the outer minimization (over a fixed $P_Z$) trivial. We now set $U=X$ in~\eqref{eq:C:star} to get
	\begin{IEEEeqnarray*}{rCl}
	C^{*,obl}&:=& \max_{P_{U|Z},x(.,.)} \min_{Q_{S}:[Q_S P_{Z}]_Z=P_Z} I(U;Y)-I(U;Z)\\
	&=& \max_{P_{X}} \min_{Q_{S}} I(X;Y),
	\end{IEEEeqnarray*}	
	where the last equality follows from $I(U;Z)=I(X;Z)=0$ as $X\perp Z$. Thus, under a \emph{state-oblivious} encoder, the randomized coding capacity $C_r^{obl}=C^{*,obl}$, and when positive, the capacity under codes with stochastic encoder $C^{obl}_s=C^{*,obl}$~(cf.~\cite[pg.~220]{csiszar-korner-book2011}).
	
		\emph{Zero or no state-myopicity:} The encoder observes $Z=S$, and hence, the outer minimization in~\eqref{eq:C:star} is now over $Q_S$. This also makes the inner minimization over $Q_S$ in~\eqref{eq:C:star} trivial. Thus, under a \emph{state-omniscient} encoder,  $C^*$ in~\eqref{eq:C:star} simplifies to
	\begin{IEEEeqnarray*}{rCl}
	C^{*,omn}&:=& \min_{Q_S} \max_{P_{U|S},x(.,.)} I(U;Y)-I(U;S).
	\end{IEEEeqnarray*}
	%where the RHS follows by replacing $Z$ with $S$ and making other appropriate changes.
	%
	This retrieves the results in~\cite{ahlswede-it1986}. In particular, the randomized coding capacity $C_r^{omn}=C^{*,omn}$, which equals the capacity under codes with  stochastic encoder $C_s^{omn}$, when $C_s^{omn}>0$.% is positive.
	
We determined the randomized coding capacity for the AVC $W_{Y|X,S}$ under a state-myopic encoder, and then showed that it  equals the capacity for codes with stochastic encoder when the latter is positive. It remains to be shown, however, how both $C_r$ and $C_s$ compare when $C_s$ equals zero. This question has been completely resolved for the two special cases discussed earlier. It is interesting to note that both these models behave quite differently. Under a state-oblivious encoder, $C_s^{obl}$ exhibits a dichotomy~\cite{ahlswede-1978,csiszar-narayan-it1988-2}, i.e., either $C^{obl}_s=C^{obl}_r$ or $C^{obl}_s=0$ (even when $C_r^{obl}>0$). This contrasts with the capacity under a state-omniscient encoder, where $C^{omn}_s=C^{omn}_r$~\cite{ahlswede-it1986}. Further, the deterministic coding capacity too has been characterized for these two special cases of our model. However, the standard approach (cf.~\cite[pg.~623]{ahlswede-it1986})  of `extracting' a `good' deterministic code, which is common to both problems, does not appear to work in our more general setting, thereby making this problem challenging. 
%It would be instructive to characterize the same for the general state-myopic encoder model.     
Other interesting extensions include studying the effects of causality and cost constraints at the encoder/adversary, as well as the generalization to continuous alphabets. Some of these are currently under investigation.
	\section*{Acknowledgment}
A.~J.~Budkuley acknowledges insightful discussions with B. K. Dey (IIT Bombay) and V. M. Prabhakaran (TIFR Mumbai) in an earlier work which proved beneficial here. Also, helpful discussions with S.~Vatedka (CUHK) are gratefully acknowledged. This work was partially funded by a grant from the University Grants  Committee of the Hong Kong Special Administrative Region (Project No.\  AoE/E-02/08) and RGC GRF grants 14208315 and 14313116.
\bibliographystyle{IEEEtran}
\balance
\bibliography{IEEEabrv,References}
\appendices
\section{Proof of Achievability}\label{app:ach:r}
We now present the detailed proof of achievability and begin with the code description.\\
\noindent \emph{Code Construction:}
\begin{itemize}
\item Our random code $\cC$ is a list of individual codes $\cC(T_Z)$, where $T_{Z}\in\cP^Q(\cZ)$. We utilize a randomized Gel'fand-Pinsker binned code for every $\cC(T_Z)$. This list of codes is shared as the common randomness $\Theta$ between the encoder-decoder \emph{a priori}.  
\item For $T_Z\in\cP^Q(\cZ)$, the codebook $\cC(T_Z)$ comprises of $2^{n R_U(T_Z)}=2^{n(R(T_Z)+\tilde{R}(T_Z))}$ vectors 
$\vec{U}_{j,k}$, where $j=1,2,\dots,2^{nR(T_Z)}$ and $k=1,2,\dots, 2^{n\tilde{R}(T_Z)}$. Thus, there are $2^{nR(T_Z)}$ bins indexed
by  $j$, with each bin containing $2^{n\tilde{R}(T_Z)}$ codewords indexed by $k$. Let $\cB^{(T_Z)}_m$ denote the bin with index $m$. We choose  $R(T_Z)< \max_{P_{U|Z},x(\cdot,\cdot)} \min_{Q_{S}\in\sQ(T_Z)} I(U;Y)-I(U;Z)$ and fix 
\begin{IEEEeqnarray*}{rCl}
(P_{U|Z},x(\cdot,\cdot))=\argmax_{P'_{U|Z},x'(\cdot,\cdot)} \min_{Q_{S}\in\sQ(T_Z)} I(U;Y)-I(U;Z)
\end{IEEEeqnarray*}
The exact choice of $R(T_Z)$, $R_{U}(T_Z)$, $\td{R}(T_Z)>0$ is specified later.  Every codeword $\vec{U}_{j,k}$ is chosen i.i.d. $\sim P_U$, where $P_U:=[P_{U|Z} T_{Z}]_{U}$. 
Thus, our code $\cC$ is the list containing $\cC(T_Z); T_Z\in\cP^Q(\cZ)$.
\end{itemize}
\emph{Encoding:}
\begin{itemize}
\item Given input message $m$ and $\bZ$, the encoder first determines the type $T_{\bZ}$ so as to identify the codebook 
$\cC(T_{\bZ})$ (as well as the corresponding optimal pair $(P_{U|Z},x(\cdot,\cdot) )$ ). Within this codebook, it finds a codeword $\bU_{m,l}$, where $m\in\{1,2,\dots, 2^{nR(T_{\bZ})}\}$ and $l=\{1,2,\dots, 2^{\tilde{R}(T_{\bZ})}\}$, such that 
\begin{equation}\label{eq:encoder:condition:dmc}
\|T_{\vec{U}_{m,l},\bZ}-P_{U|Z} T_{\bZ}\|_{\infty}\leq \delta_2(\delta).
\end{equation}
Here $\delta_2(\delta)>0$ is a fixed constant (we state the choice of 
$\delta_2(\delta)$ later in Claim~\ref{claim:cover})\footnote{Here $\delta>0$ is a function of $\epsilon$, such that $\delta\rightarrow 0$ as $\epsilon \rightarrow 0$.}. 
This condition implies that $(\vec{U}_{m,l},\bZ)$ are jointly
typical according to the distribution $P_{U|Z} T_{\bZ}$. If the encoder is unable to find such
$\vec{U}_{m,l}$, then it chooses $\bU_{1,1}$. If more than
one $\vec{U}_{m,l}$ satisfying~\eqref{eq:encoder:condition:dmc} exist, then the
encoder chooses one uniformly at random from amongst them. Let
$\vec{U}=\vec{U}_{M,L}$ denote the chosen codeword. 
\item The encoder then transmits $\bX$, where $X_i=x(U_i, Z_i)$, $i=1,2\dots, n$, and $T_{\bZ}$ to the decoder. 
\end{itemize}
\emph{Decoding:}
\begin{itemize}
\item Let the decoder receive the  channel output $\by$ and the type $T_{\bz}$.  It identifies the codebook $\cC(T_{\bz})$ chosen by the encoder.  
\item For some fixed parameter $\gamma(\delta)>0$
(the choice of $\gamma(\delta)$ is indicated later in Lemma~\ref{lem:dec:events}), the decoder first identifies the set of
codewords 
\begin{IEEEeqnarray*}{rCl}\label{eq:dec:cond}
\mathcal{L}_{\gamma(\delta)}(\by)=\Big\{ \vec{u}\in \mathcal{B}^{(T_{\bz})}&: \|T_{\vec{u},\vec{y}}-[P_{U|Z} T_{\bz} \vec{1}_{\{X=x(u,z)\}}  T_{S}  W_{Y|X,S} ]_{U,Y}\|_{\infty}\leq\gamma(\delta), \text{ some } T_{S}\in \sQ(T_{\bz}) \Big\}, 
\end{IEEEeqnarray*}
where $\sQ(T_{\bz})=\{T_{S}\in\sT^n(\cS): [T_{S} P_{Z|S}]_Z\stackrel{f(\epsilon)}{\approx} T_{\bz} \}$.
%and $f(\epsilon)>0$ and $f(\epsilon)\rightarrow 0$ as $\epsilon\rightarrow 0$. Here $T^1_X\stackrel{f(\epsilon)}{\approx} T^2_X$ implies $\|T_X^1-T_X^2\|_\infty \leq f(\epsilon)$.
%
\end{itemize}
\noindent \emph{Probability of error analysis}\\
Let the chosen codeword be $\bU=\bU_{M,L}$. A decoding error occurs if one or more of the following events occur.
\begin{IEEEeqnarray*}{rCl}
E_{enc}&=& \{ (\bU_{j,k}, \bZ) \not\in  \cT^n_{\delta}(P_{U|Z} T_{\bZ}),\,\forall j,k\}\\
E_{dec1}&=& \{ \bU \not\in  \cL_{\gamma(\delta)}(\bY)\}\\
E_{dec2}&=& \{ \bU_{m',l'} \in  \cL_{\gamma(\delta)}(\bY) \text{ for some } m'\neq m, k'\}.
\end{IEEEeqnarray*}
Now from the union bound, the probability of decoding error is given by
\begin{IEEEeqnarray}{rCl}\label{eq:P:E}
\bbP(E)\leq \bbP(E_{enc})+\bbP(E_{dec,1}|E^c_{enc})+\bbP(E_{dec,2}|E^c_{enc}).
\end{IEEEeqnarray}
We will show that for every $\epsilon>0$ there exists small enough $\delta>0$ 
such that $\bbP(E)\rightarrow 0$ as $n\rightarrow \infty$.
We now make the following claims. Let $\delta_0<<\delta$. 
%\ajb{AJB: express exactly in terms of $\delta$ if possible.}
%
\begin{claim}
Given any  $\bs\in\cS^n$,  $\bs\in T^n_{\delta_0}(T_{\bs})$.
\end{claim}
\begin{IEEEproof}
Note that $\bs\in\cT^n_{\delta'}(T_\bs)$,   for any $\delta'>0$. The result follows by choosing $\delta'=\delta_0$.
\end{IEEEproof}
\begin{claim}[~\cite{bdp-arxiv2017}]\label{claim:cond:typ}
Let $\bs\in\cT^n_{\delta_0}(T_{\bs})$ and let $\bZ$ be generated via the DMC $P_{Z|S}$ under input $\bs$. Then, %
\begin{IEEEeqnarray*}{rCl}
\bbP((\bs,\bZ)\in\cT^n_{3\delta_0}(T_{\bs}P_{Z|S}))\geq 1-|\cS| |\cZ|e^{-2n\delta_0	}.\yesnumber \label{eq:s:z:expo}
\end{IEEEeqnarray*}
\end{claim}
%The proof of this result is along the lines of the conditional typicality lemma~\cite[pg.~]{elgamal-kim}.
%
%\ajb{AJB: this proof only for completeness. It is the \emph{conditional typicality lemma} from~\cite{elgamal-kim} re-derived for this notion of strong typicality.}
\begin{IEEEproof}
The proof of this result appears in~\cite{bdp-arxiv2017}, and is given here for completeness. 
We intend to show that
\begin{align*}
&\mathbb{P} \left(\left|T_{\bs,\bZ}(s,z)-P_S(s) P_{Z|S}(z|s) \right| > 3\delta_0\right) 
\end{align*}
is exponentially small for all $s,z$.
We consider two cases. \\
{\it Case I:} $T_{\bs}(s)\leq \dnot$.
As $\bs\in\cT^n_{\dnot}(P_S)$, this implies that 
$ P_S(s)\leq T_{\bs}(s)+\dnot \leq 2\delta_0.  $
Then,~$\forall (s,z)$,
\begin{IEEEeqnarray*}{rCl}
\left|T_{\bs,\bZ}(s,z)-P_S(s) P_{Z|S}(z|s) \right| &=&\left|T_{\bs}(s)T_{\bZ|\bs}(z|s)-P_S(s) P_{Z|S}(z|s) \right|\\
&\leq& \max \left( T_{\bs}(s)T_{\bZ|\bs}(z|s), P_S(s)P_{Z|S}(z|s)\right)  \\
&\stackrel{}{\leq}& 2\dnot \cdot 1  \\
&=&2\dnot.
\end{IEEEeqnarray*}
Thus, for such $s$, $\bbP \left(\left|T_{\bs,\bZ}(s,Z)-P_S(s) P_{Z|S}(z|s) \right| > 2\dnot\right)=0$.\\
{\it Case II:} $T_{\bs}(s)> \dnot$.
Using Chernoff-Hoeffding's  theorem~\cite[Theorem~1]{hoeffding-1963} for each $z\in\cZ$, we have
\begin{IEEEeqnarray*}{rCl}
\mathbb{P}(|P_{Z|S}(z|s)-T_{\bZ|\bs}(z|s)| >\dnot, \text{ for any } z ) \leq |\cZ|e^{-2n \delta^3_0 }.
\end{IEEEeqnarray*}
Now, it can be easily checked that 
$|P_{Z|S}(z|s)-T_{\bZ|\bs}(z|s)| \leq \dnot$ and $|P(s)-T_{\bs}(s)| \leq
\dnot$ together imply 
$$\left|T_{\bs}(s)T_{\bZ|\bs}(z|s)-P_S(s) P_{Z|S}(z|s) \right| \leq 2\dnot+\dnot^2 \leq 3\dnot.$$
Hence,~\eqref{eq:s:z:expo} follows by taking union bound over all $s\in\cS$.
\end{IEEEproof}
To proceed, let us define the ``good'' event $A_{S,Z}:=\{(\bS,\bZ)\in\cT^n_{3\delta_0}(T_{\bS}P_{Z|S})\}$.
\begin{claim}
Conditioned on the event $A_{S,Z}$, $\bZ$ is $\delta_1$-typical w.r.t. $P_{Z}=[T_{\bS}P_{Z|S}]_Z$ where $\delta_1:=3|\cS|\delta_0\rightarrow 0$ as $\delta\rightarrow 0$. Equivalently, $\|T_{\bZ}-P_Z\|_{\infty}\leq \delta_1$.
\end{claim}
The proof of this claim is straightforward.
%\ajb{AJB: above claim basically uses that fact that if $(\bx,\by)\in\cT^n_{\delta}$, then $\bx\in\cT^n_{\delta |\cY|}$.}
\begin{claim}
Let $\bU$ be generated i.i.d. via distribution $P_{U}$. Then, it follows that with probability at least $1-|\cU|e^{-2n\delta^2}$, $\bU\in\cT^n_{\delta}(P_{U})$. 
\end{claim}
%
%\ajb{AJB: this is $\delta(\epsilon)$ chosen (i.e., an appropriate code!)}
The proof of this result is straightforward, and follows via the Chernoff bound.
To proceed, let us define the event $A_U:=\{\bU\in\cT^n_{\delta}(P_U)\}$.
\begin{claim}\label{claim:cover}
Conditioned on the event $A_{U}$, there exists $\delta_2(\delta)>0$ and $f_2(\delta,\epsilon)>0$, where $\delta_2(\delta),f_2(\delta,\epsilon)\rightarrow 0$ as $\delta,\epsilon\rightarrow 0$ such that encoding is successful and the encoder finds a codeword $\bU$ with probability at least $1-2^{-nf_2(\delta,\epsilon)}$ such that $(\bU,\bZ)\in\cT^n_{\delta_2}(P_{U|Z}T_{\bZ})$. 
\end{claim}
The proof of this claim uses the covering lemma~\cite{elgamal-kim}. This result specifies the parameter $\delta_2(\delta)$ which appears in the specification of the encoder, and implies that $\bbP(E_{enc})\rightarrow 0$ as $n\rightarrow \infty$.
We define the ``good'' event $A_{U,Z}:=\{(\bU,\bZ)\in\cT^n_{\delta_2}(P_{U|Z}T_{\bZ}\}$.
%\\
%\ajb{AJB: we have $(\bU,\bZ)\in\cT^n_{\delta_2}(P_{U|Z} T_{\bz})$, and we show $(\bU,\bZ)\in\cT^n_{\delta_3}(P_{U|Z} P_Z)$ with $P_Z$ instead of $T_\bZ$.} 
%
\begin{claim}\label{claim:z:u:typ}
Under $E^c_{enc}$ and $A_{U,Z}$, $(\bU,\bZ)$ are jointly $\delta_3$-typical according to the distribution 
$P_Z P_{U|Z}$, where 
$P_Z=[T_{\bS} W_{Z|S}]_Z$ and
$\delta_3(\delta):=3|\cS|\dnot(\delta) + \delta_2(\delta) \rightarrow 0$ as $\delta \rightarrow 0$.
\end{claim}
\begin{IEEEproof}
Note that
\begin{align*}
\|P_ZP_{U|Z} - T_{\bU\bZ}\|_{\infty} & \leq \|P_ZP_{U|Z} - T_{\bZ}P_{U|Z}\|_{\infty}
               + \|T_{\bZ}P_{U|Z} - T_{\bU\bZ}\|_{\infty}\\
& \leq 3|\cS|\dnot + \delta_2 \quad \text{(using $A_{UZ}$ and $E^c_{enc}$)}\\
& = \delta_3,
\end{align*}
where $\delta_3=3|\cS|\dnot + \delta_2$.
\end{IEEEproof}
%
%\ajb{AJB: we have $(\bS,\bZ)$ and $(\bU,\bZ)$ jointly typical separately. Our aim is to use the Markov lemma to conclude that $(\bS,\bZ,\bU)$ are together jointly typical. THe next lemma is towards establishing that $\bU$ satisfies the conditions in the refined ML.}
\begin{claim}[\cite{bdp-arxiv2017}]\label{claim:unif:dist}
There exists $g(\delta)>0$, where $g(\delta)\rightarrow 0$ as $\delta\rightarrow 0$, such that $\forall \bu\in\cT^n_{\delta_3}(P_{U|Z}P_Z|\bz)$,
\begin{IEEEeqnarray*}{rCl}\label{eq:tdX:dist}
P_{\bU}(\bU =\bu|\bZ=\bz)\leq 2^{-n(H(U|Z) -g(\delta) )},
\end{IEEEeqnarray*}
where $H(U|Z)$ is computed with the distribution $P_{U|Z}P_Z$.
\end{claim}
\begin{IEEEproof}[Proof of claim]
The proof of this result is along the lines in~\cite{bdp-isit2017}.
We have two cases.\\
{\it Case 1:} When $\bu\in \cT^n_{\delta_2}(P_{U|Z}T_{\bz}|\bz) \bigcap \cT^n_{\delta_3}(P_{U|Z}P_Z|\bz)$.
Then we note that 
\begin{IEEEeqnarray*}{rCl}
&\mathbb{P} &\left( \bU =\bu|\bZ=\bz \right) \\
&\stackrel{}{=}& \mathbb{P} \left( \bU=\bu,\bU \in\cT^n_{\delta_3}(P_{U|Z}T_{\bz}|\vec{z})\big|\bZ=\bz  \right)\\
&=&\mathbb{P} \left( \bU \in\cT^n_{\delta_3}(P_{U|Z}T_{\bz}|\vec{z})\big|\bZ=\bz\right)\mathbb{P} \left( \bU=\bu\big|\bZ=\bz, \bU \in\cT^n_{\delta_3}(P_{U|Z}T_{\bz}|\vec{z}) \right)\\
&\leq& \mathbb{P} \left( \bU=\bu\big|\bZ=\bz, \bU \in\cT^n_{\delta_3}(P_{U|Z}T_{\bz}|\vec{z}) \right)\\
&=& \mathbb{P} \left( \bU_{1,1}=\bu\big|\bZ=\bz, \bU_{1,1} \in\cT^n_{\delta_3}(P_{U|Z}T_{\bz}|\vec{z}) \right)\\
&=& \frac{\mathbb{P} \left( \bU_{1,1}=\bu|\bZ=\bz\right)}{ \mathbb{P}\left(\bU_{1,1} \in\cT^n_{\delta_3}(P_{U|Z}T_{\bz}|\vec{z})|\bZ=\bz \right)}\\
&\stackrel{}{\leq}& 2\cdot \mathbb{P} \left( \bU_{1,1}=\bu|\bZ=\bz\right)
\quad \text{(since  $\mathbb{P}\left(\bU_{1,1} \in\cT^n_{\delta_3}(P_{U|Z}T_{\bz}|\vec{z})|\bZ=\bz \right) \rightarrow 1$ as $n\rightarrow \infty$)}\\
&\stackrel{}{\leq}& 2^{-n(H_{P_{U|Z}T_{\bz}}(U|z) -g_1(\delta_3) )}
\end{IEEEeqnarray*}
where $g_1(\delta_3)\rightarrow 0$ as $\delta_3\rightarrow 0$.
Since $||P_Z - T_{\bz}||_1 \leq |\cZ|\cdot ||P_Z - T_{\bz}||_\infty
\leq 3|\cS|\delta_0$, and
$||P_ZP_{U|Z} - T_{\bz}P_{U|Z}||_1 \leq |\cU||\cZ|\delta_3$,
using \cite[Lemma~2.7]{csiszar-korner-book2011}, we get
\begin{IEEEeqnarray*}{rCl}
& |H_{P_Z}(Z) - H_{T_{\bz}}(Z)| \leq 3|\cS|\dnot\cdot\log\left(
\frac{1}{3|\cS|\dnot}\right)\\  
& |H_{P_ZP_{U|Z}}(U,Z) - H_{T_{\bz}P_{U|Z}}(U,Z)| \leq |\cU||\cZ|\delta_3\cdot
\log\left( \frac{1}{\delta_3}\right)  
\end{IEEEeqnarray*}
Together, the above two equations imply
\begin{align*}
& |H_{P_ZP_{U|Z}}(U|Z) - H_{T_{\bz}P_{U|Z}}(U|Z)| \leq 2|\cU||\cZ|\delta_3\cdot
\log\left( \frac{1}{\delta_3}\right). \label{eq:H:u:z}
\end{align*}
By defining $g_2(\delta_3) := g_1(\delta_3) + 2|\cU||\cZ|\delta_3\cdot
\log\left( \frac{1}{\delta_3}\right)$, we get
\begin{align*}
\mathbb{P} \left( \bU =\bu|\bZ=\bz \right)
& \stackrel{}{\leq} \frac{1}{2}\cdot 2^{-n(H_{P_{U|Z}P_{Z}}(U|Z) -g_2(\delta_3) )}.
\end{align*}
{\it Case II:} 
When $\bu\not\in \cT^n_{\delta_2}(P_{U|Z}T_{\bz}|\bz) $. For such a $\bu$,
the encoder outputs it only if $\bU_{1,1}= \bu$ and there is no codeword
which is jointly typical with $\bz$ w.r.t. $P_{U|Z}T_{\bz}$.
Thus,
\begin{align*}
\bbP\left( \bU =\bu|\bZ=\bz \right)
& \stackrel{}{\leq} \bbP\left( \bU_{1,1} =\bu|\bZ=\bz \right)\\
& \leq 2^{-n(H_{[P_{U|Z}T_{\bz}]_U}(U) - g_3(\delta_2))}\\
& \leq 2^{-n(H_{[P_{U|Z}P_Z]_U}(U) - g_4(\delta_2))},
\end{align*}
where $g_4(\delta_2) = g_3(\delta_2) + |\cU|^2|\cZ|\delta_3\cdot \log\left(
\frac{1}{|\cU||\cZ|\delta_3}\right)$.

Combining the two cases, and taking $g(\delta) = \max (g_2(\delta_3),
g_4(\delta_3))$, the lemma follows.
\end{IEEEproof}
We now state a refined version of the Markov lemma from~\cite{bdp-it2017}\footnote{See also~\cite[Lemma~14]{bdp-arxiv2017}.
\removed{
In the refined Markov lemma presented in~\cite{bdp-it2017}, condition (b) also has a lower
bound on $P_{\bZ}(\bz)$. However, the lower bound is not used in the proof
given in \cite{bdp-it2017}, and hence, can be removed. Here, we state this lemma without any lower
bound. We also note that condition (a) and the upper
bound on the probability of a typical sequence imply that probability
of too many typical sequences can not be too small; and so some essence
of the lower bound in condition (b) is already implied by these. 
Thus, it is not surprising that the lower bound is not needed for the 
lemma to hold.}}.
\begin{lemma}[Refined Markov Lemma~\cite{bdp-it2017}]\label{lem:ref:markov:lemma}
Suppose $X\rightarrow Y\rightarrow Z$ is a Markov chain, i.e., $P_{X,Y,Z}=P_{Y}P_{X|Y} P_{Z|Y} $. Let $(\vec{x},\vec{y})\in \mathcal{T}^n_{\dnot}\left(P_{X,Y}\right)$ and $\vec{Z} \sim P_{\vec{Z}}$ be such that
\begin{enumerate}[(a)]
\item $\mathbb{P}\left((\vec{y},\vec{Z})\not\in
\mathcal{T}^n_{\dnot}\left(P_{Y,Z}\right)\right)\leq \epsilon$, where $\epsilon>0$,
%$\epsilon_n\rightarrow 0$ as $n\rightarrow \infty$.
%, where $P_{Y,Z}^{(n)}=P_{Y}^{(n)}P_{Z|Y}$, and

\item for every $\vec{z}\in \mathcal{T}^n_{\dnot}\left(P_{Y,Z}|\vec{y}\right)$,
\begin{equation*}\label{eq:condition:markov}
P_{\vec{Z}}(\vec{z})\leq 2^{-n(H(Z|Y)-g(\dnot))},
\end{equation*}
for some $g:\mathbb{R}^{+}\rightarrow\mathbb{R}^{+}$, where $g(\dnot)\rightarrow 0$ as $\delta_0\rightarrow 0$.
\end{enumerate}
Then, there exists $\delta:\mathbb{R}^{+}\rightarrow\mathbb{R}^{+}$, where $\delta (\dnot) 
\rightarrow 0$ as $\dnot \rightarrow 0$, such that
\begin{equation*}
\mathbb{P}\left((\vec{x},\vec{y},\vec{Z})\not \in \mathcal{T}^n_{\delta(\dnot)}\left(P_{X,Y,Z}\right)   \right)\leq 2|\cX||\cY||\cZ| e^{-n K} +\epsilon.
\end{equation*}
Here $K>0$ and $K$ does not depend on $n$, $P_{X,Y}$, $P_{\vec{Z}}$ or $(\vec{x},\vec{y})$ but does depend on $\dnot$, $g$ and $P_{Z|Y}$. Further, the $\delta$ function does not depend on $(\bx,\by)$, $P_{X,Y}$ or $P_{\bZ}$.
%\ajblue{AJB: All references to dependence on $P_{Z|Y}$ and not on $P^{\min}_{Z|Y}$. }
%$P^{\min}_{Z|Y}:=\min_{(y,z) : P_{Z|Y}(z|y)>0} P_{Z|Y}(z|y)$.  
\end{lemma}
Let us now use the above lemma to prove the following result.
\begin{claim}\label{claim:s:z:u:typ}
There exists $\dfourd>0$, where $\dfourd\rightarrow 0$ as
$\delta\rightarrow 0$, such that, except for a small
probability, $(\bS,\bZ,\bU)$
is jointly $\dfour$-typical w.r.t. $T_{\bS} P_{Z|S}P_{U|Z}$.
\end{claim}
\begin{IEEEproof}
We assume that  $A_{X,J,Y,Z}$ is true, and use the refined Markov lemma (Lemma~\ref{lem:ref:markov:lemma}) on the Markov chain $S \rightarrow Z \rightarrow U$. Now through Claims~\ref{claim:z:u:typ} and~\ref{claim:unif:dist}, we know that $\bU$ is chosen such that both
conditions (a) and (b) in Lemma~\ref{lem:ref:markov:lemma} are satisfied.  This establishes the claim.
\end{IEEEproof}
Let us define the resulting ``good'' event as $A_{S,Z,U}:=\{(\bS,\bZ,\bU)
\in \cT^n_{\dfour}(T_{\bS} P_{Z|S} P_{U|Z}\}$.
\begin{claim}\label{claim:s:z:u:x:typ}
Conditioned on the event $A_{S,Z,U}$ and given that $\bX$ be generated via the distribution $\vec{1}_{\{X=x(U,Z)\}}$, then with probability at least $1-|\cS| |\cZ| |\cU||\cX|e^{-2n\delta_4	}$,
%\begin{IEEEeqnarray*}{rCl}
%\bbP((\bs,\bZ)\in\cT^n_{3\delta_0}(T_{\bs}P_{Z|S}))\geq 1-|\cS| |\cZ|e^{-2n\delta_0	}.\yesnumber \label{eq:s:z:expo}
%\end{IEEEeqnarray*}
%
$(\bS,\bZ,\bU,\bX)$
is jointly $\dfive$-typical w.r.t. the distribution $T_{\bS} P_{Z|S}P_{U|Z}\vec{1}_{\{X=x(U,Z)\}}$, where $\dfive:=3\dfour$.
\end{claim}
\begin{IEEEproof}
The proof of this follow from Claim~\ref{claim:cond:typ}.
\end{IEEEproof}
We define the resulting ``good'' event as $A_{S,Z,U,X}:=\{(\bS,\bZ,\bU,\bX)
\in \cT^n_{3\dfour}(T_{\bS} P_{Z|S} P_{U|Z}\vec{1}_{\{X=x(U,Z)\}}\}$.
\begin{claim}\label{claim:s:z:u:x:y:typ}
Conditioned on the event $A_{S,Z,U,X}$ and given that $\bY$ be generated via the distribution $W_{Y|X,S}$, then with probability at least $1-|\cS| |\cZ| |\cU||\cX||\cY|e^{-2n\delta_4	}$,
%\begin{IEEEeqnarray*}{rCl}
%\bbP((\bs,\bZ)\in\cT^n_{3\delta_0}(T_{\bs}P_{Z|S}))\geq 1-|\cS| |\cZ|e^{-2n\delta_0	}.\yesnumber \label{eq:s:z:expo}
%\end{IEEEeqnarray*}
%
$(\bS,\bZ,\bU,\bX,\bY)$
is jointly $3\dfive$-typical w.r.t. the distribution $T_{\bS} P_{Z|S}P_{U|Z}\vec{1}_{\{X=x(U,Z)\}} W_{Y|X,S}$.
\end{claim}
\begin{IEEEproof}
As in the earlier claim, the proof follows from Claim~\ref{claim:cond:typ}.
\end{IEEEproof}
We define this ``good'' event by $A_{S,Z,U,X,Y}:=\{(\bS,\bZ,\bU,\bX,\bY)
\in \cT^n_{3\dfour}(T_{\bS} P_{Z|S} P_{U|Z}\vec{1}_{\{X=x(U,Z)\}} W_{Y|X,S}\}$.
We now state the result which handles the other two terms which appear in the expression for $\bbP(E)$ in~\eqref{eq:P:E}.
\begin{lemma}\label{lem:dec:events}
Let the chosen codeword be $\bU$, where $\bU\in\cT^n_{\delta}(P_U)$, and let the observed channel output be $\bY$. Then, 
\begin{itemize}
\item there exists $\gamma(\delta)>0$, where $\gamma(\delta)\rightarrow 0$ as $\delta\rightarrow 0$, such that $\bU\in\cL_{\gamma(\delta)}(\bY)$, except for an exponentially small probability.
\item there exists $f_3(\delta,\epsilon)>0$, where $f_3(\delta,\epsilon)\rightarrow 0$ as $\delta,\epsilon\rightarrow 0$, such that 
\begin{IEEEeqnarray*}{rCl}
\bbP(\bU_{m',l'} \in\cL_{\gamma(\delta)} (\bY) \text{ for some } m'\neq m, l')\leq 2^{-nf_3(\delta,\epsilon)}.
\end{IEEEeqnarray*}
\end{itemize}
\end{lemma}
\begin{IEEEproof}
The first part of the proof follows from the following claim.
\begin{claim}
There exists $\gamma(\delta)>0$, where $\gamma(\delta)\rightarrow 0$ as $\delta\rightarrow 0$, such that except for an exponentially small probability, $\bU\in \cL_{\gamma(\delta)}(\bY)$. 
\end{claim}
\begin{IEEEproof}
Consider the event $A_{S,Z,U,X,Y}$. Under this event, $(\bU,\bY)$ are
$\gammad$-typical w.r.t. the distribution $P_{U,Y}=[T_{\bS} P_{Z|S} P_{U|Z}\vec{1}_{\{X=x(U,Z)\}} W_{Y|X,S}]_{U,Y}$.
Thus, the claim now follows from Claim~\ref{claim:s:z:u:x:y:typ}.
\end{IEEEproof}
%
%This claim specifies the $\gamma(\delta)$ parameter which appears in the definition of the decoder in~\eqref{eq:dec:cond}. 
The following claim establishes the second part.
\begin{claim}
There exists $f_2(\delta,\epsilon)>0$, where $f_2(\delta,\epsilon)\rightarrow 0$ as $\delta,\epsilon\rightarrow 0$, such that
\begin{IEEEeqnarray*}{rCl}
\mathbb{P}\left(\vec{U}_{M',L'}\in \mathcal{L}_{\gamma(\delta)}(\bY), \text{ for some } M'\neq M,L' \right)\leq 2^{-n f_2(\delta,\epsilon)}.
\end{IEEEeqnarray*}
\end{claim}
\begin{IEEEproof}
We note that the codewords $\{\bU_{M',L'}\}_{M'\neq M}$ are independently generated. This implies that the codewords $\{\bU_{M',L'}\}_{M'\neq M}$ and $\bY$ are independent. Let us fix a conditional type $T_{S}\in \sQ(T_{\bZ})$ at the decoder, and let the resulting distribution $P_{U,Y}=[T_S P_{Z|S} P_{U|Z} \vec{1}_{\{X=x(U,Z)\}} W_{Y|X,S} ]_{U,Y}$. Then, we have
\begin{align*}
\mathbb{P}\big(\exists m',l':m'\neq m:(\vec{U}_{m',l'},\bY)\in \cT^n_{\gamma(\delta)} (P_{U,Y})) \leq  2^{-n\td(f_2)}
\end{align*}
for some $\td{\ftwo} \rightarrow 0$ as $\delta, \epsilon \rightarrow 0$. This follows from the packing lemma~\cite[Lemma~3.1]{elgamal-kim}. 
By taking the union bound over all conditional types $T_{S}\in\sQ(T_{\bZ})$ (note that there are at most polynomial number of types in $n$), we get
\begin{IEEEeqnarray*}{rCl}
\mathbb{P}\big(\exists m'\neq m, l': (\vec{U}_{m',l'},\bY)\in \cT^n_{\gamma(\delta)} (P_{U,Y}) \text{  for some } T_{S}\in \sQ(T_{\bZ})) &\leq& (n+1)^{|\cU||\cY|} 2^{-n\tdftwo}\\
&\leq & 2^{-n\ftwo}.
\end{IEEEeqnarray*}
The claim is, thus, established.
\end{IEEEproof}
As both the parts have been proved, this completes the proof of the lemma.
\end{IEEEproof}
This lemma provides the $\gamma(\delta)$ parameter used in the specification of the decoder. Further, it implies that both the terms appearing in the $\bbP(E)$ expression, viz., $\bbP(E_{dec1}|E^c_{enc}), \bbP(E_{dec2}|E^c_{enc})\rightarrow 0$ as $n\rightarrow \infty$.

As each of the terms in the RHS of~\eqref{eq:P:E} are vanishing as $n\rightarrow \infty$, it follows that $\bbP(E)\rightarrow 0$ as $n\rightarrow \infty$. Thus, we have shown that for every $\epsilon>0$, there exists a $\delta>0$ such that $\bbP(E)\rightarrow 0$ as $n\rightarrow \infty$. This completes the proof of achievability.

\end{document}